\documentclass[journal]{IEEEtran}

\ifCLASSINFOpdf

\else

\fi

\usepackage{rotating}

\usepackage{lscape}
\usepackage{booktabs}
\usepackage{lscape}
\usepackage{colortbl}
\usepackage{multirow}
\usepackage{amssymb}

\usepackage{pdflscape} 
\usepackage{multirow}
\usepackage{supertabular}

\usepackage[english]{babel}
\usepackage[utf8]{inputenc}
\usepackage{amsmath}
\usepackage{graphicx}
\usepackage[colorinlistoftodos]{todonotes}
\usepackage{silence}
\usepackage{hyperref}
\usepackage{url}

\usepackage{breakurl}
\usepackage{siunitx} 
\usepackage{booktabs} 

\usepackage{graphics}
\usepackage{adjustbox}
\usepackage{amsfonts}
\usepackage{rotating}

\usepackage{lscape}
\usepackage{booktabs}
\usepackage{lscape}
\usepackage{colortbl}
\usepackage{multirow}
\usepackage[T1]{fontenc}
\usepackage{multirow, makecell}
\usepackage{subfig}
\usepackage{longtable}

\newcolumntype{L}{>{\raggedright\arraybackslash}p}

\usepackage{mathtools}

\DeclarePairedDelimiter\abs{\lvert}{\rvert}%

\makeatletter
\let\oldabs\abs
\def\abs{\@ifstar{\oldabs}{\oldabs*}}
\usepackage{mathtools}

\newcommand\cc{C\&C}

\newcommand\dns[1]{{\small \texttt{#1}}}
\newcommand\dlabel\dns
\newcommand\shell\dns

\usepackage{tikz}

\usepackage[ruled,vlined]{algorithm2e}

\usepackage{multirow, makecell}

\usepackage{lipsum}

\usepackage{listings}
\usepackage{xcolor}
\begin{document}

%
\title{Investigation of Advanced Persistent Threats Network-based Tactics, Techniques and Procedures }
%
%
%

\author{\IEEEauthorblockN{Almuthanna Alageel
and
Sergio Maffeis}\\

\IEEEauthorblockA{Department of Computing \\
Imperial College London \\
London, United Kingdom}
}

\thispagestyle{plain}
\pagestyle{plain}
\maketitle

\begin{abstract}

The scarcity of data and the high complexity of Advanced Persistent Threats (APTs) attacks have created challenges in comprehending their behavior and hindered the exploration of effective detection techniques.
To create an effective APT detection strategy, it is important to examine the Tactics, Techniques, and Procedures (TTPs) that have been reported by the industry. These TTPs can be difficult to classify as either malicious or legitimate. When developing an approach for the next generation of network intrusion detection systems (NIDS), it is necessary to take into account the specific context of the attack explained in this paper.

In this study, we select 33 APT campaigns based on the fair distribution over the past 22 years to observe the evolution of APTs over time. We focus on their evasion techniques and how they stay undetected for months or years. 
We found that APTs cannot continue their operations without \cc\ servers, which are mostly addressed by Domain Name System (DNS). 
We identify several TTPs used for DNS, such as Dynamic DNS, typosquatting, and TLD squatting. 
The next step for APT operators is to start communicating with a victim. We found that the most popular protocol to deploy evasion techniques is using HTTP(S) with 81\% of APT campaigns. HTTP(S) can evade firewall filtering and pose as legitimate web-based traffic. DNS protocol is also widely used by 45\% of APTs for DNS resolution and tunneling. We identify and analyze the TTPs associated with using HTTP(S) based on real artifacts. 
We also investigated the excessive use of fallback channels to evade NIDS that rely on volume-based features. We identify that 60.6\% of APT campaigns split their traffic over multiple IP addresses. We also highlight other TTPs that can also be equipped during a campaign, such as non-application protocols or data obfuscation, which are frequently found in our analysis. In conclusion, we outline a roadmap for the future development of next-generation APT detection systems. Our study highlights the necessity of innovative solutions to counter evolving attack patterns, improve threat visibility, and reinforce proactive defense strategies.  

\end{abstract}


\IEEEpeerreviewmaketitle

\section{Introduction}

According to the National Institute of Standards and Technology (NIST) special publication (SP) 800-61 \cite{cichonski2012computer}, an APT can be defined as a highly-skilled adversary who targets organizations with specified objectives with multiple attack vectors over a long period. 
APTs cost the UK government and worldwide nearly GBP 27 billion annually and USD 1 trillion \cite{tankard2011advanced}, respectively. The majority of objectives of APTs are espionage, data exfiltration and damaging critical infrastructure. These campaigns' well-trained cyber security personnel come from military and governmental organizations, academia, and R\&D entities \cite{apt1}. 
To create an effective APT detection strategy, it is important to examine the Tactics, Techniques, and Procedures (TTPs) that have been reported by the industry. These TTPs can be difficult to classify as either malicious or legitimate because they can cause significant overlap in features between APTs and legitimate connections \cite{alageel2022earlycrow}. Therefore, when developing an approach for the next generation of Network Intrusion Detection Systems (NIDS), it is necessary to take into account the specific context of the attack explained in the following sections.

Several survey papers explain the impact of APT operations, and discuss related work. 
Taleb et al. \cite{talib2022apt} discuss the techniques used for APT beaconing detection in the literature review. Our paper investigates the techniques used from the adversary's point of view to evade the detection techniques. 
Lemay et al. \cite{lemay2018survey} is the first paper to identify public reports for APTs. They categorized the technical references by threat actor, content, and type. Based on these reports, they describe 38 APT campaigns across different regions. Our paper is inspired by \cite{lemay2018survey} for starting the investigation of APTs based on the technical reports. However, our profiling focuses on the techniques used for \cc\ to evade APT detection. Based on our datasets, we also provide some evidence for such techniques to facilitate the next generation of APT detection.

Stojanovi et al. \cite{stojanovic2020apt} discuss the models used for the APT lifecycle. They also identify network and system datasets for APT. Our paper considers only major models, and then we use them as a taxonomy for our profiling. We collect our dataset specifically for APT campaigns related to \cc\ artifacts, as we will discuss below. 
Alshamrani et al. \cite{alshamrani2019survey} present a general attack tree model in addition to several case studies for different objectives, such as stealing data or damaging critical infrastructure.  %
In this paper, our main aim is to understand APT behavior based on technical reports released in the past, in addition to our technical analysis based on our collected datasets. There are more than a hundred APT campaigns, according to MITRE CO.  \cite{strom2017finding}. We track those with high impact over 22 years. Our selection is also based on the trusted resources where they are available to provide a reliable analysis. Our aim is to facilitate the next generation of NIDS to capture these kinds of threats that not only rely on a single technique but a planned arsenal of TTPs equipped with their malware and operations.

This paper aims to investigate the evasion techniques and how the APT can remain undetected by NIDS for a long time. We analyze APT campaigns using 118 technical reports and our own investigation. We find that spearphishing links have been consistently popular for delivering backdoors over the past 22 years. This highlights the need for a detective approach to defend against malicious domains. Spearphishing attachments are also prevalent and can bypass current host-based intrusion detection.
Our research examines whether APTs consist of a single malware or a diverse arsenal. We confirm that APTs utilize multiple malware families, with APT 1 alone using up to 26 different families. Many of these backdoors communicate with remote command and control servers. Detecting this malicious traffic can help identify if an organization is under an APT campaign attack. A large majority (84.8\%) of APT campaigns use backdoors, while 78.7\% utilize RATs, and only 6\% involve botnets. This highlights the heavy reliance on customized backdoors and RATs by APTs, with botnets being rarely used.

Since our paper focuses on defending against APTs using network data, we concentrate on our findings related to popular protocols and TTPs used for communication with command and control servers.
We observe that APT campaigns have consistently employed HTTP since 2001 to evade NIDS detection. It is crucial to carefully detect malicious connections using the HTTP protocol, although not every single connection will use it. Additionally, 81\% of APT campaigns utilize HTTPS to bypass NIDS relying on HTTP plaintext features. DNS protocols are used by 45\% of APTs, while others use HTTP with preconfigured IP addresses.
Furthermore, 24.2\% of APT campaigns distribute malware through spearphishing links, necessitating the consideration of detecting malicious URLs. SMTP, P2P, SOCKS5, SMB, and FTP follow as the next commonly used protocols.

To develop a reliable approach for detecting APTs, we study their use of TTPs to bypass NIDS. APTs frequently employ data obfuscation through protocol impersonation, as shown in Figure \ref{fig:profiling_results}.c with the example of Mivast and Skula malware impersonating legitimate HTTPS. The fallback channel, used by 60.6\% of APT campaigns, divides traffic volume among multiple command and control servers. This technique has become increasingly popular over time to evade volume-based detection approaches.
Multi-level encrypted channels, used by 54.5\% of APT campaigns, protect against decryption if TLS traffic is blocked. Domain fronting and multi-hop proxy are exploited by more than half (51.5\%) of APT campaigns, while 24.2\% solely use domain fronting to conceal the location of the remote server. Other DNS-based TTPs include dynamic DNS (27.2\%) and exploitation of DGA or DNS tunneling (24.2\%). This emphasizes the importance of detecting malicious domains and UDP-based traffic.
%
%

The structure of this paper can be described as follows: First, we explain the damage caused by APTs in the past in Section \ref{sec:back}. Based on that, we define the methodology that we follow for our investigation and the datasets we used, in Section \ref{sec:meth}. We also discuss the APT lifecycle based on the literature in Section \ref{sec:lifecycle}. To study the malicious TTPs accurately, we start with identifying the relevant information from the network perspective from technical reports in Section \ref{Section: Analysis of APT}. We review 118 reports released by the industry since 2001. We select 33 APT campaigns based on a fair distribution over the past 22 years to observe the evolution of APT over time. We describe many TTPs relevant to network security. Next, those TTPs that are not detailed in the industry reports from the traffic behavior perspective. We resort to our datasets (Section \ref{sec:meth}) to provide details on their behavior and how they evade network defenses in Sections \ref{sec:dnsttps} and \ref{sec:trafficttps}. Finally, we discuss our findings in Section \ref{sec:diss}.

Our main findings can be summarized as follows:
\begin{itemize}
    \item APT Campaign Analysis: The study examines 33 APT campaigns based on 118 technical reports and real investigations, shedding light on their characteristics and behaviors. It specifically focuses on the prevalent use of spearphishing links and attachments as a means of delivering backdoors, emphasizing the need for effective defense against malicious domains. 
    \item Malware Diversity in APTs: The paper investigates whether APTs rely on a single type of malware or employ a variety. By reporting the backdoors used in each campaign, we confirm that APTs utilize a collection of malware, with up to 26 different families employed. It reveals that APTs primarily rely on customized backdoors and RATs, while botnets are seldom utilized.
    \item Protocol Analysis: The paper discusses the prevalent use of HTTP and HTTPS protocols by APT campaigns. It highlights the evasion tactics employed, such as using HTTP to bypass NIDS and relying on HTTPS to bypass NIDS relying on HTTP plaintext features. The study also explores the use of DNS protocols and alternative protocols like SMTP, P2P, SOCKS5, SMB, and FTP.
    \item TTP Analysis: The research focuses on TTPs employed by APT campaigns. It examines data obfuscation through protocol impersonation and the utilization of fallback channels, multi-level encryption, domain fronting, and multi-hop proxies. The paper also highlights DNS-based TTPs like dynamic DNS and DGA or DNS tunneling.

\end{itemize}

\section{Historical Perspective} \label{sec:back}
Since 2006, a large-scale cyber-espionage APT campaign called APT 1, a.k.a Comment Crew, has been compromising more than 140 organizations in twenty different sectors, including state administrations, energy, health, scientific research entities, aerospace, satellites, telecommunication, IT, and financial services \cite{apt1}. More than ninety organizations had failed to detect the malicious activity of APT 1 for an average of a year, while other organizations had been challenged to detect over 40 malware families used by APT 1 for up to four years and ten months. 87\% of these organizations are headquartered in English-speaking countries with 6.5 TB of compressed data exfiltrated for more than ten months from a single organization without detection. The espionage operation collected business plans, senior management emails, system design, simulation technologies, and user credentials. From 2011 to 2013, APT1 expanded \cc\ servers to establish more than 937 with 988 FQDNs resolved with unique IP addresses and to transmit its traffic over HTTPS with thirteen different X.509 certificates \cite{apt1}.

From the beginning of 2007 until September 2014, APT 28, a.k.a Fancy Bear, scanned 1.7 million vulnerable IPs in Ukraine alone against several sectors such as government entities, telecommunication, and aerospace companies \cite{apt28b}. The same campaign was famous for espionage against the Georgian military and other Eastern European states. The campaign had been successfully hidden from detection by mimicking legitimate domains, frequently updating to open C\&C multichannel using DGA for its Fall-back Channels \cite{apt28c}.

In mid-2009, the Operation Aurora campaign targeted more than 34 organizations, including Yahoo, Symantec, Morgan Stanely, and Google, that could not detect the infiltration for over six months \cite{tankard2011advanced}. Another campaign called APT 32, a.k.a OceanLotus or Operation Cobalt Kitty, delivered fileless payloads, i.e. in-memory, to hide their traces from forensics engineers \cite{apt32c}. The operation transferred 46 binary files and 24 scripts via several C\&C channels, including Denis and Goopy Trojan, PowerShell script and outlook macro backdoors,  Cobalt Strike, Don't-Kill-My-Cat evasion tool, custom NetCat and IP tools \cite{apt32a}. The stealthy C\&C channel operated over DNS tunneling to evade both NIDS and Firewalls \cite{apt32c}.

Undetected for over seven years \cite{apt29b}, a well-resourced cyberespionage group called APT 29, a.k.a CozyDuke, was found to infiltrate the US White House successfully, in addition to defense, energy, and financial institutions in Western Europe and China \cite{apt29a} at the end of 2015. APT 29 is known for deploying many malware toolsets, most of which belong to the Duke malware family \cite{apt29b}. APT 3, a.k.a Buckeye, has been targeting nearly 84 organizations, including government departments in the US, the UK, and Hong Kong, in low and slow mode since 2009 \cite{apt3b}. The campaign employed zero-day exploits through spearphishing to drop a remote access Trojan called Pirpi \cite{apt3b}. APT 3 is known to employ a vast arsenal of malware and TTP, including Remote Access Tools (RAT), backdoors, keyloggers and Lazagne to extract passwords from the current application and exfiltrate all data back to C\&C server \cite{apt3b}.

Another stealthy and extremely successful campaign~\cite{apt18b} is called Wekby or APT 18. The operation started in 2009 and has been known to use DNS as a medium to communicate with C\&C servers via HTTPBrowser, gh0st RAT, and Pisloader for over six years without detection \cite{apt18a}. Another cyber espionage, APT 37, utilized many zero-day vulnerabilities through spearphishing links or attachments, strategic web compromise (SWC), and Torrent file-sharing from 2012 until 2018 in support of military activities of the state against South Korea, Japan, and the Middle East \cite{apt37}. The campaign delivered different malware families, including SLOWDRIFT, KARAE and POORAIM malware and TTP toolsets to exfiltrate highly classified data to cloud services and then to wipe the master boot record (MBR) of the organization's data center \cite{apt37}.

\section{Methodology} \label{sec:meth}
Several researchers argue whether APT is a special case of a multi-stage attack or another malware family. According to Khattak et al. \cite{khattak2013taxonomy}, Aurora was a specialized botnet with a cyber espionage objective against Google. Zhao et al. \cite{zhao2015detecting} define one primary difference that makes APT malware different from Botnet. While the former's purpose is cyber espionage or data exfiltration, the latter is destruction, such as denial of service attacks. On the other hand,  Vormayr et al. \cite{vormayr2017botnet} describe Blackenergy, Duqu 2.0, and Regin as Botnet and APT alongside Conficker and Phatbot with a minor difference in whether the attack is targeted or not.

Ussath et al. \cite{ussath2016advanced} confirm that APT is a campaign and present a survey of APT campaigns based on 22 reports in terms of initial compromise methods, their lateral movement techniques and \cc\ protocols used to transmit the \cc\ payloads. However, details on \cc\ protocols and other settings are not available, as we will discuss in this paper. Authors in \cite{chen2014study} provide an analysis of 4 campaigns with respect to each stage of the Cyber Kill Chain model, without focusing on persistent and evasion properties, a wide range of protocols and a variety of channels. 
This leads us to limit our analysis to the network level of an APT campaign to illustrate how persistent, evasive and stealthy APT traffic is, and how we can design a robust network-based intrusion detection accordingly.

Last but not least, the primary source of information comes from industry, due to the relative monopoly on information related to APT campaigns. According to Lemay et al., \cite{lemay2018survey}, there is no alternative to industrial resources on this topic because targeted organizations request forensics services from industrial labs, who might publish this confidential information upon the client's permission.

Our investigation in this paper relies on published works by scholars and covers those reports released by practitioners in major security vendors. In addition, we confirm our analysis with the dataset we collected to identify and explain TTPs based on the artifacts. 

\begin{figure}[h]
\centering
\includegraphics[scale=0.6]{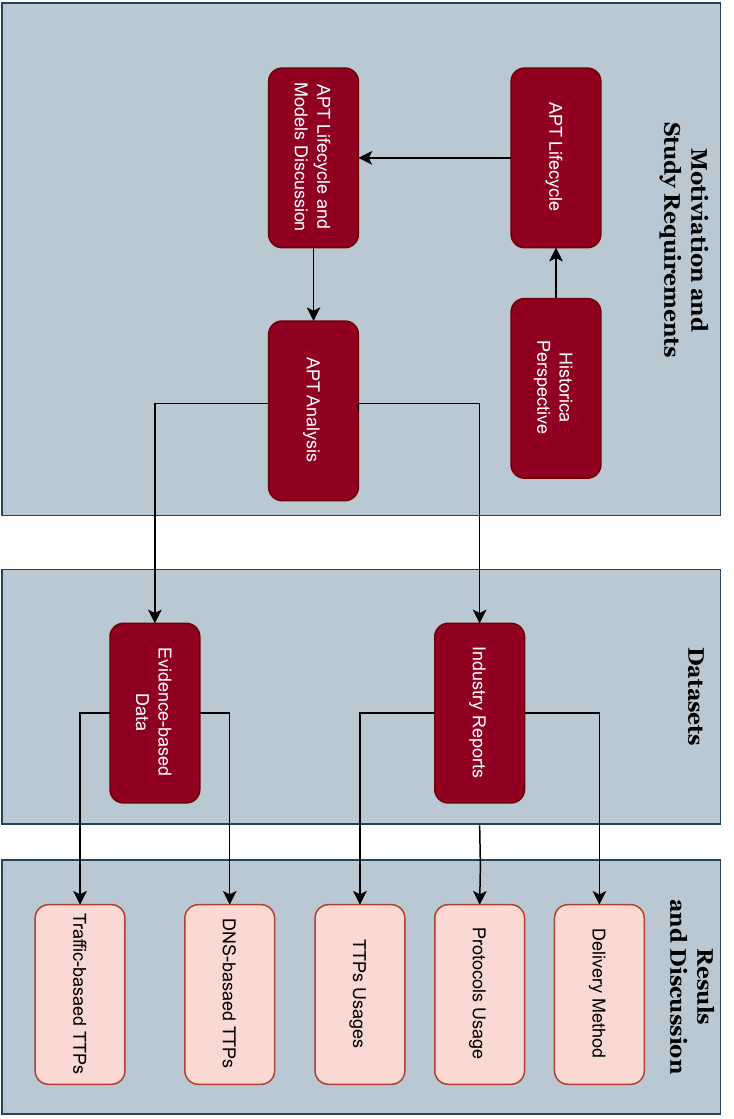}
\caption{Methodology.} \label{figure:meth}
\end{figure}

The datasets we collected and used in this paper are summarized as follows:
\begin{itemize}

    \item Industry Reports Public Information (\textbf{IRPI}): In this dataset, we collect the related information for APTs across 33 campaigns for the last 22 years. We selected these campaigns based on two factors: the fair distribution for the evolution of APTs across 22 years. Second, the availability of data from industry vendors. For the latter one, when we collect data for several campaigns for the same timeframe, we 
    choose the one that causes major damage, which, consequently, produces several public industry reports. 

    \item APT, Phishing and Legitimate Domains (\textbf{HEALP}) \cite{alageel2021hawkeye}: Based on 118 reports, we collect xxx domains. We also track their historical WHOIS and DNS resource records using SecurityTrail. We aim to identify the common characteristics of APT campaigns in different regions. Then, we aim to analyze the techniques that have shifted over 22 years so we can find the weakest spot in our defenses and expect the next potential attacks. 

    \item APT, Botnets and Legitimate traffics (\textbf{APTrace}) \cite{alageel2022earlycrow}: This traffic is based on HTTP(S) connections that are used by live APT, botnets, and legitimate traffic. 

    \item Malware Capture Facility Project (\textbf{MCFP}) based on the complication of \cite{alageel2022earlycrow}:

    \item APT, Botnets and Legitimate traffic (\textbf{APTracePluse}): in this dataset, we extend the dataset based on \cite{alageel2022earlycrow} to include protocols used beyond HTTP(S) such as Raw TCP and UDP, which are popular as we found in Section \ref{Section: Analysis of APT}.

\end{itemize}

\section{APT Lifecycle} \label{sec:lifecycle}
Understanding how an APT campaign behaves from the first day until the mission is accomplished can be crucial in identifying the attack surface at each stage and provide deep insight for our profiling. This also helps in extracting the relative information for \cc\ stage and TTPs used to evade detection. Lockheed Martin Cyber Kill Chain$^\text{{\tiny{\textregistered}}}$ is considered as the first model that was able to describe APT proposed in 2011 \cite{hutchins2011intelligence}. Cyber Kill Chain$^{\text{\tiny{\textregistered}}}$ is then followed by Mandiant and MITRE ATT\&CK, both of which are increasingly accepted among the research community. However, there are two further models proposed by Vries et al. \cite{de2012systems} and Attack Pyramid by Giura and Wang \cite{giura2012context}, each of which describes an APT as a normal attack lifecycle without any APT propriety, therefore, we omit these two models from our discussion for the brevity.

\subsection{Lockheed Martin Cyber Kill Chain$^{\text{\tiny{\textregistered}}}$}
\begin{figure}[h]
\centering
\includegraphics[scale=0.33]{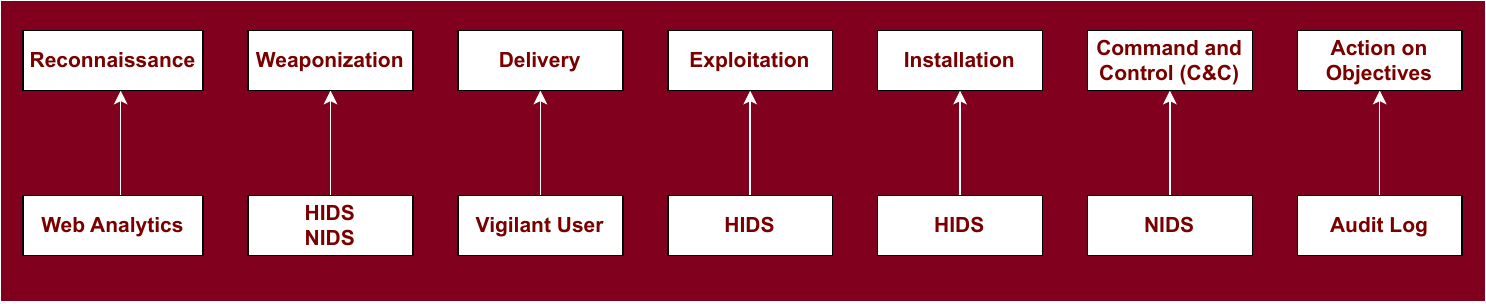}
\caption{Lockheed Martin Cyber Kill Chain detective controls \cite{hutchins2011intelligence}.} \label{figure:LM}
\end{figure}

Hutchins et al. propose Cyber Kill Chain \cite{hutchins2011intelligence}, which is part of a broader Intelligent Driven Defense framework funded and adopted by Lockheed Martin \cite{hutchins2011intelligence}. The model is widely adopted by several papers, including   \cite{chen2014study, yadav2015technical, kiwia2018cyber,ioannou2013markov}. It describes APTs in seven phases in one direction for one time, as depicted in Figure \ref{figure:LM}.

Cyber Kill Chain starts with the reconnaissance stage, which includes passive reconnaissance, scanning and enumeration. Then, the adversary starts developing malware and coupling it with exploits in the weaponisation stage. The malicious payload is then attached to a medium at the delivery stage, whether that medium could be a malicious link to a website that hosts the malicious payload or attached with a file, e.g. PDF or MS Word. When a target is vulnerable to that exploitation (stage 4), an installer component attached to malware tends to establish a backdoor at stage five and be able to communicate back with the threat actor through the \cc\ server in stage six. The last stage is where the action can be made, which varies according to the APT objectives from the beginning, including espionage, data exfiltration and disruption. Hutchins et al. recommend a matrix of security measures against each stage, including detection, prevention, disruption, degrading, and deceiving, with respect to stage order. We illustrate in Figure \ref{figure:LM} a course of action in terms of detective controls recommended by the authors  \cite{hutchins2011intelligence}.

\subsection{Mandiant Attack Model}
Mandiant Attack Model is proposed by FireEye to analyze APT 1 \cite{apt1}. The model is adopted by many researchers, including \cite{milajerdi2018holmes,mathew2010situation}. This section will highlight the similarities and differences between the Mandiant and Cyber Kill Chain Models.

The Mandiant model begins with initial reconnaissance, which matches the first stage on the Cyber Kill Chain. Then, a threat actor initially compromises a host in stage 2, which matches stages 3-5 in the Cyber Kill Chain, i.e. delivery, exploitation and installation. It is worth noting that Mandiant does not consider weaponisation as well as merging three stages from the Cyber kill chain into one stage. This is because Mandiant focuses on defense strategy more than considering the full picture from the adversary's perspective. Next, the adversary establishes a foothold once a backdoor is already installed, which is equivalent to \cc\ - stage six - in Cyber Kill Chain. 

\begin{figure}[h]
\centering
\includegraphics[scale=0.58]{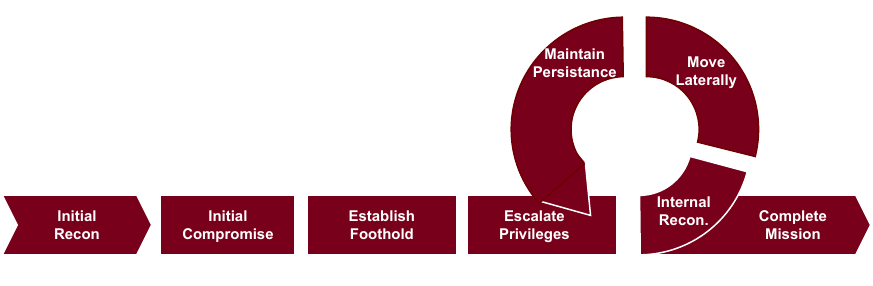}
\caption{Mandiant attack model \cite{apt1}.} \label{figure:Mandiant}
\end{figure}

The next four stages are performed $n$ times by the adversary until an organization takes defensive action upon detecting the malicious activities. The first stage of the $n$ cycle stages is privilege escalation, where an adversary either delivers another zero-day exploit or passes the hash to take over a root account, for instance. Next, the adversary uses built-in operating system commands or a custom one to understand the local environment intellectually. The adversary moves around the network as a privileged and legitimate user who can evade all detective controls. The adversary can then launch an OS scheduler, a persistent technique in the following stage, to install further backdoors or to exfiltrate data according to a specific schedule in the final stage as depicted in Figure \ref{figure:Mandiant}. The iterative cycle might be conducted multiple times in order to open fall-back channels and maintain APT persistence.

\subsection{ATT\&CK \textsuperscript{TM} Model}

\begin{figure}[h]
\centering
\includegraphics[width=8.5cm,height=7cm]{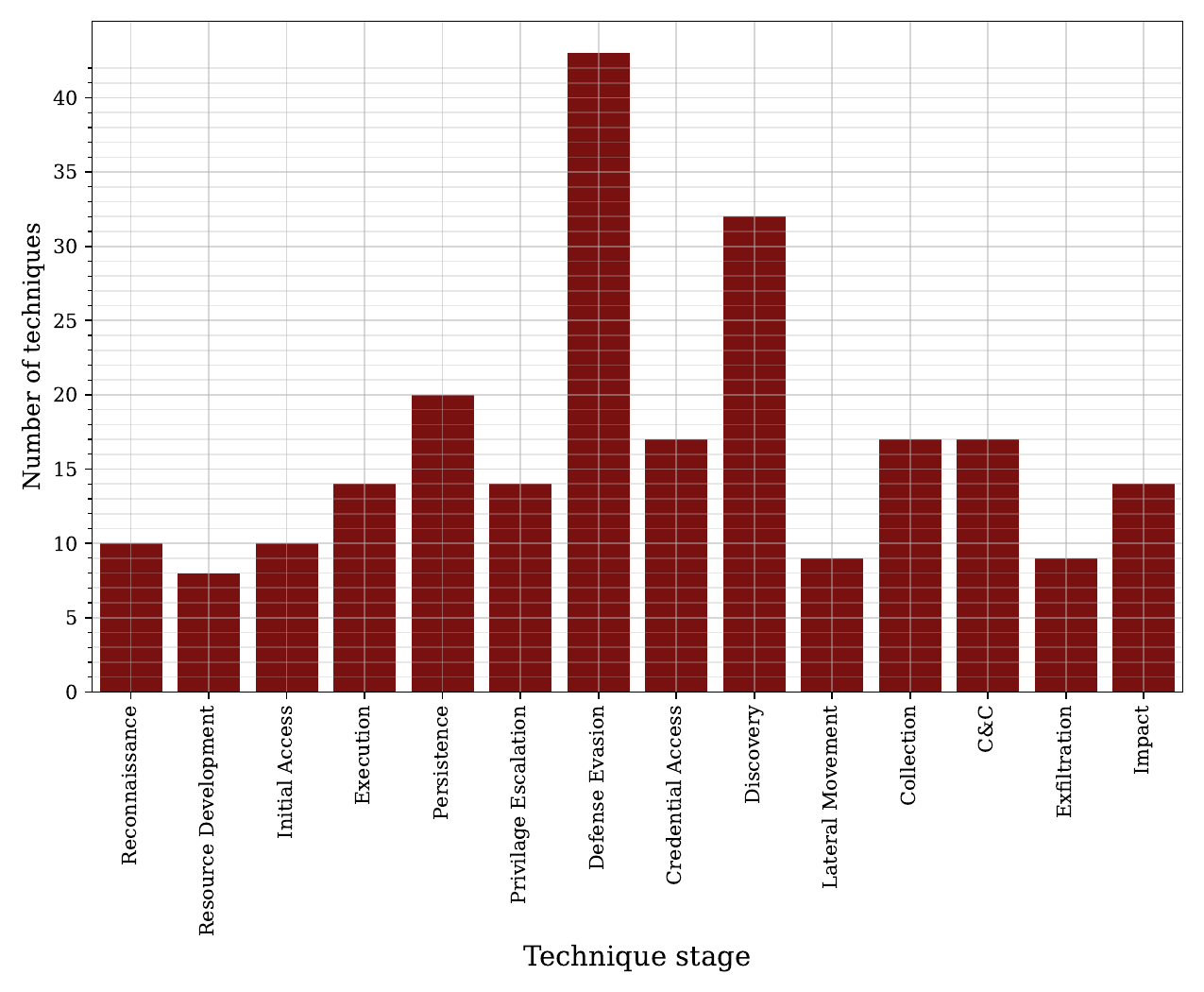}
\caption{MITRE ATT\&CK techniques per stage.} \label{figure:ATTCK}
\end{figure}

While Cyber Kill Chain and Mandiant APT attack models are presented in an abstract view and are focused on the lifecycle of an APT campaign, MITRE ATT\&CK presents a matrix for enterprise to describe Tactics, Techniques and Procedures (TTP) of an APT campaign. The matrix and its TTPs are based on the observation of tens APT campaigns \cite{strom2017finding}. The TTP are presented across twelve stages as depicted in Figure \ref{figure:ATTCK}. The total number of TTPs is 130, which an APT can select from such a pool. For example, APT 1 used 22 TTPs across all stages. It is worth paying attention to the new stages not presented by other models, such as defense evasion, credential access, discovery, and collection. The defense evasion stage covers evasion techniques from the hardware to the application level. However, our analysis in Section \ref{Section: Analysis of APT} leads to an expanded view of network evasion techniques based on our observations. Internal reconnaissance in other models is divided here into discovery and collection, where the former represents an early stage of internal reconnaissance and the latter refers to the late one where accounts, files, and emails, for example, are collected and sent back to \cc\ servers in \cc\ stage.

As we can notice from the description above and shown in Figure \ref{figure:ATTCK}, the matrix presents the TTP with respect to each stage, but the stages are not chained to form a lifecycle. Therefore, ATT\&CK Matrix might be used as a companion and a pool of TTPs to other lifecycle models to overcome the absence of subsequent behavior of an APT in order to provide proper security measures. 

\subsection{Discussion on APT Models} \label{APTModels-Disscussion}

We note that Cyber Kill Chain takes the full picture from the adversary's point of view. For instance, stages 1-4 are out of the target's network perimeter. Establishing a foothold on the model is represented in stage six, while multiple subsequent actions are conducted and not presented well. Furthermore, unlike Mandiant, the cyber kill chain does not represent the cyclic nature of APT, where an APT starts expanding inside the target's network. In addition, the Cyber kill Chain merges the internal reconnaissance and lateral movement stages of the Mandiant model under the action on the objective stage. 

Mandiant fits for research that focuses on host intrusion detection and endpoint protection more than the network perspective. This is because the Mandiant attack model does not explicitly describe \cc\ stage, as we can see from Figure \ref{figure:Mandiant}, which is our scope in this paper. As regards to ATT\&CK Matrix, it does adopt a lifecycle scheme. Instead, a pool of TTPs is presented in a concrete way and lacks abstraction for future TTP.

It is important to include in the APT models how their custom malware operates once it infects the victim. APT malware are those malicious tools known to be used by APT campaigns. They usually do not participate in DDoS attacks, sending spam or propagating to other hosts to spread infections \cite{farinholt2020dark}. APT malware is classified as targeted malware with different functions to communicate with \cc. For instance, a RAT is typically composed of a builder, stub, and controllers. The builder initiates a new instance stub upon the infection. The stub stays on the victim with preconfigured FQDN or IP to communicate to RAT's controller placed at the \cc\ server \cite{rezaeirad2018schrodinger}. Trojans, spyware, and keyloggers may also be composed to connect \cc\ at a low profile. However, some malware, such as DarkCommet, includes these functions in one ecosystem  \cite{farinholt2020dark}, which may capture the audio, explore files and drop malicious tools through visiting URLs \cite{farinholt2017catch}. Griffon, used by FIN 7, can gather information, load Meterpreter, and take screenshots \cite{griffon19}. 

Once APT actors drop malware on infected hosts, they may be inactive instantly. Several malicious activities can be done automatically before the APT actors control the victim. For instance, malware connects automatically to the \cc\ server, and other automatic operations are followed, such as information gathering, dropping more malicious tools or payloads, and establishing fallback channels. Therefore, it is crucial to detect these malicious tools initially, whether the actors start controlling them or not, as long as they are connected to \cc\ servers. 

Based on this discussion, we summarize the requirements for developing an updated APT campaign network-based attack model:
\begin{itemize}
    \item \cc\ stage should be included in the proposed APT life cycle to identify when \cc\ is needed for an adversary and why.
    \item \cc\ stage should be divided into multiple stages to provide the details of attack behavior. Stages may include DNS resolution, locating \cc\ servers and malicious \cc\ traffic.
    \item Evasion techniques at \cc\ stages reported by MITRE ATT\&CK need to be considered based on real evidence, such as dynamic resolution, data obfuscation and fallback channels.
    \item The type of APT malware operations should be pointed out to connect the gaps between the multiple stages of \cc. 
\end{itemize}

\section{Analysis of APT Campaigns} \label{Section: Analysis of APT}

In this section, we present our survey of 33 APT campaigns presented in Tables \ref{tab:long_table1} and \ref{tab:long_table2}, and we provide an analysis of these campaigns from the network perspective. First, we focus on the variant names for the same campaign, followed by their points of entry and the delivery methods. Next, we explore some evasion techniques for most campaigns. Then, we describe what payloads are exchanged between the victims and \cc\ server and how they are obfuscated. All these characteristics are provided based on IoC, public reports, and evidence materials. Sources include, among others, Bitdefender, ESET, Trend Micro, Sophos, McAfee, F-Secure Lab, Symantec, FireEye Mandiant, Kaspersky Lab, Checkpoint and evidence from the Department of Homeland Security in the USA.

At the beginning of this paper, we briefly define the APT and start arguing whether APT refers to a special type of malware or a well-resourced campaign. After the survey presented in Tables \ref{tab:long_table1} and \ref{tab:long_table2}, we have seen how a single APT campaign is equipped with multiple malware, including RAT, backdoors or trojans. As expected, only APT 35 and Sandworm use botnets out of 33 APT campaigns. From now on, we refer to the malware used by APT for \cc\ communications with confirmed IP addresses as APT malware unless the industry reports it as a botnet. 
Table \ref{tab:long_table2} shows the persistence and stealthiness properties of APT malware, which increase the chances of evading defenses over a long period of time. 
%

There are remarkable differences between an APT and other malware, such as botnets or ransomware, that use \cc\ communication. 
For instance, when APTs initially compromise a target and drop a backdoor to communicate with \cc\ servers, they tend to implant other RATs and further backdoors to continue their operation with other \cc\ servers. 
In practice, once a SOC engineer discovers a backdoor or a list of IoCs belonging to an APT, an incident response plan will be carried out to contain the incident. However, with the persistence property, an APT tends to implant several backdoors against the same organization, each of which exploits different zero-day vulnerabilities \cite{apt1}. 

For instance, Elderwood APT launch their attack through zero-day exploits of CVE-2012-0779, CVE-2012-1875, and CVE-2012-1889,  followed by another five zero-day exploits in one month against the same target in order to deliver backdoor Trojans \cite{aptElderwoodA} and that is a sign of highly intelligent and knowledgeable engineers who in charge of these campaigns with unlimited resources. In addition, we notice another example of how APT differs from typical malware or attacks represented by CozyDuke, a.k.a APT 29. The campaign uses a modular malware platform that includes the command execution module, password stealer, NTLM hash stealer, downloader, loaders and backdoors such as HammerDuke OnionDuke, CosmicDuke, and SeaDuke.
As a result, the campaign might continue without detection and complete espionage and disruption objectives. Nevertheless, the APT campaign carefully develops its own custom malware, FQDNs and IP addresses and is able to exfiltrate over time to stay in stealthy mode against the same target.

Next, we focus on the network characteristics of the APT campaigns covered in the previous section.
Table \ref{table:2.2} summarizes the APT campaign in terms of their adaptation of protocols that carry out \cc\ traffic, the tool type of \cc, and the most frequent evasion techniques adopted for APT campaigns. We notice that most APT campaigns use DNS and HTTP(S) for their operations through their attack period. It does not mean every malware uses the two protocols, but many frequently use them.


\begin{landscape}
  \begin{center}
    {\footnotesize
    \setlength\tabcolsep{1pt}

    \tablefirsthead{%
      \toprule
      \makecell{\multirow{1}{*}{\textbf{\makecell{APT Campaign}}}} 
        & \makecell{\multirow{1}{*}{\textbf{\makecell{Delivery Method}}}}  
        & \makecell{\multirow{1}{*}{\textbf{\makecell{Vulnerability}}}} 
        & \makecell{\multirow{1}{*}{\textbf{\makecell{Backdoors}}}} \\
      \midrule
      \midrule
    }

    \tablehead{%
      \multicolumn{4}{l}{\tablename\ \thetable\ -- (Continued)} \\
      \toprule
      \makecell{\multirow{1}{*}{\textbf{\makecell{APT Campaign}}}} 
        & \makecell{\multirow{1}{*}{\textbf{\makecell{Delivery Method}}}}  
        & \makecell{\multirow{1}{*}{\textbf{\makecell{Vulnerability}}}} 
        & \makecell{\multirow{1}{*}{\textbf{\makecell{Backdoors}}}} \\
      \midrule
      \midrule
    }

    \tabletail{%
      \bottomrule
      \multicolumn{4}{r}{\textit{Continued on the next page}} \\
    }

    \tablelasttail{%
      \bottomrule
    }

    \topcaption{APT Campaigns From Network Perspective - Part I.}
    \label{tab:long_table1}

    \begin{supertabular}{L{3cm}*{1}{L{4cm}}*{1}{L{6cm}}*{1}{L{10cm}}}

  APT 1, a.k.a Comment Crew & Spearphishing attachment \cite{apt1}. & N/A & Standard Backdoors: Poison Ivy, Gh0st  Beachhead Backdoors: Beachhead Family i.e. WEBC2 (>16 variants) e.g. WEBC2-QBP AURIGA, BANGAT, BISCUIT, BOUNCER, CALENDAR, COMBOS, COOKIEBAG, DAIRY, GLOOXMAIL, GOGGLES, GREENCAT, HACKSFASE, HELAUTO, KURTON, LONGRUN, MACROMAIL, MANITSME, MINIASP, NEWSREELS, SEASALT, STARSYPOUND and SWORD \cite{apt1}.\\
  \\
      
       APT 2, a.k.a Putter Panda & Spearphishing attachment abc.scr Dropper To install 4H RAT \cite{apt2a}. & N/A & 3PARA RAT, 4H RAT, httpclient, pngdowner \cite{apt2a}.  \\
       \\
      APT 3, a.k.a Gothic Panda a.k.a Pirpi  Buckeye & Spearphishing attachment malicious RAR  Spearphishing Link Browser Exploit \cite{apt3e}. & For initial access: Flash SWF file CVE-2015-3113 $\dagger$.  For privilege escalation: CVE-2014-4113 $\dagger$ \cite{apt3a}. Others: CVE-2015-5119 $\dagger$ CVE-2010-3962 $\dagger$ CVE-2014-1776 $\dagger$ CVE-2014-6332 \cite{apt3d}.  & Backdoor.Pirpi (12 variants) \cite{apt3b}, SHOTPUT, Backdoor.APT.CookieCutter and PlugX \cite{apt3e}. 
        \\
      APT 10 \\ a.k.a menuPass & Spearphishing to deliver EvilGrab or ChChes \cite{apt10a}. &N/A &  HTRAN, ZXProxy, ZXPortMap, PlugX, PoisonIvy, QuasarRAT, RedLeaves \cite{apt10b}. \\
        \\
      APT 12 & Spearphishing Attachment \cite{apt12}.  & MS Word CVE-2012-0158 \cite{apt12}. &RIPTIDE , HIGHTIDE \cite{apt12}, THREBYTE and WATERSPOUT.  \\
      \\
      APT 15, a.k.a Ke3chang  Vixen Panda & Spearphishing attachment/URL \cite{apt15a}. IE injection technique used by HTTP-based, Backdoors (BS2005 Malware) \cite{apt15a}. & MS Word CVE-2010-3333, Adobe PDF Reader CVE-2010-2883 \cite{apt15a}. & RoyalCli, BS2005 and RoyalDNS \cite{apt15b}.\\
      \\
      APT 16, a.k.a EPS & Spearphishing Attachment MS Word \cite{apt16b} and URL (DOORJAMB.Tools, IRONHALO or ELMER \cite{apt16a}. &  MS Word: CVE-2015-2545 $\dagger$, Windows: CVE-2015-2546 $\dagger$ \cite{apt16a}, Windows privilege escalation CVE-2015-1701 \cite{apt16b}. & IRONHALO ELMER with two variants \cite{apt16b}. \\
      \\
      APT 17 a.k.a Deputy Dog & Spearphishing URL or Social Networks (Cleaver)\cite{apt17}.  & N/A & BLACKCOFFEE \cite{apt17}. \\		
\\
      APT 18, a.k.a Wekby, Dynamite Panda& Spearphising  Xyligan RAT \cite{apt18b}.   &N/A& Xyligan.rat, gh0st.rat, HcdLoader.rat, Pisloader \cite{apt18a,apt18b}. \\
      \\
      APT 19 a.k.a. Codoso & Waterhole \cite{apt19} Spearphishing Rich Text Format (RTF) macro-enabled MS Excel \cite{apt19b}. & Windows CVE 2017-0199 \cite{apt19b}. & N/A  \\
\\
         APT 27 a.k.a Emissary Panda, Threat Group-3390, LuckyMouse & Spearphishing attachment (ZIP archive) \cite{apt27a}. & CVE-2014-6324 \cite{apt27d}, CVE-2017-11882 \cite{apt27b}. For privilege escalation: Jave SWCs CVE-2011-3544, JBoss CVE-2010-0738 \cite{apt27a}. & HyperBro a.k.a Backdoor.Win32.HyperBro (Trojan.backdoor.rat),  Trojan.Win32.Generic \cite{apt27b}, PlugX \cite{apt27a},  (Trojan.backdoor.rat), HttpBrowser (Trojan.backdoor.rat) \cite{apt27b}.  \\\\

		\bottomrule 
    \end{supertabular}
    }
  \end{center}
\end{landscape}

\begin{landscape}
  \begin{center}
    {\footnotesize
    \setlength\tabcolsep{1pt}

    \tablefirsthead{%
      \toprule
      \makecell{\multirow{1}{*}{\textbf{\makecell{APT Campaign}}}} 
        & \makecell{\multirow{1}{*}{\textbf{\makecell{Delivery Method}}}}  
        & \makecell{\multirow{1}{*}{\textbf{\makecell{Vulnerability}}}} 
        & \makecell{\multirow{1}{*}{\textbf{\makecell{Backdoors}}}} \\
      \midrule
      \midrule
    }

    \tablehead{%
      \multicolumn{4}{l}{\tablename\ \thetable\ -- (Continued)} \\
      \toprule
      \makecell{\multirow{1}{*}{\textbf{\makecell{APT Campaign}}}} 
        & \makecell{\multirow{1}{*}{\textbf{\makecell{Delivery Method}}}}  
        & \makecell{\multirow{1}{*}{\textbf{\makecell{Vulnerability}}}} 
        & \makecell{\multirow{1}{*}{\textbf{\makecell{Backdoors}}}} \\
      \midrule
      \midrule
    }

    \tabletail{%
      \bottomrule
      \multicolumn{4}{r}{\textit{Continued on the next page}} \\
    }

    \tablelasttail{%
      \bottomrule
    }


    \begin{supertabular}{L{3cm}*{1}{L{4cm}}*{1}{L{6cm}}*{1}{L{10cm}}}

 APT 28 a.k.a Sednit Sofacy Fancy Bear &  Spearphishing URL-shortener HIDEDRV rootkit with Downdelph backdoor \cite{apt28a, apt28c}. & Java CVE-2015-2590 $\dagger$, Flash CVE-2015-3043 $\dagger$, CVE-2015-5119 $\dagger$, CVE-2015-7645, Word CVE-2015-1641 $\dagger$, CVE-2015-2424 $\dagger$ \cite{apt28d}, IE CVE-2014-4076 \cite{apt28a}. For privilege escalation:  CVE-2015-2387 $\dagger$, CVE-2015-1701 $\dagger$ \cite{apt28d}. & Downdelph, CHOPSTICK, CORESHELL, Komplex, Zebrocy, JHUHUGIT and Sofacy \cite{apt28e, apt28f}. \\
      \\
      APT 29 a.k.a CozyBear CozyDuke & Spearphishing attachment and link \cite{apt29c}. & PDF Acrobat: CVE-2013-2729, MS Office:  CVE-2009-3129, CVE-2015-2424, CVE-2015-1641, CVE-2010-3333, CVE-2014-1761, CVE-2012-0158, Adobe Flash CVE-2016-4117 CVE-2016-7855 and for Privilage Escalation: CVE-2016-7255 \cite{apt29c}.  & CosmicDuke, CloudDuke, HammerDuke, SeaDuke, SeaDaddy, PinchDuke, OnionDuke, MiniDuke, HAMMERTOSS \cite{apt29b} and Cobalt Strike \cite{apt29e}.  \\\\
    
     APT 30 & Spearphishing attachment to drop BACKSPACE \cite{apt30}. & N/A & FLASHFLOOD, NETEAGLE, SHIPSHAPE, BACKSPACE SPACESHIP \cite{apt30}.  \\
\\

      APT 32, a.k.a OceanLotus & Spearphishing attachment (RTF Document) \cite{apt32b}.  & CVE-2017-11882. For privilege escalation: CVE-2016-7255 \cite{apt32b} & Cobalt Strike, Denis \cite{apt32a}, KOMPROGO, OSX OCEANLOTUS.D, PHOREAL, SOUNDBITE, WINDSHIELD \cite{apt32e}.  	   \\
      \\
      APT 33, a.k.a Elfin  Shamoon & Spearphishing Link \cite{apt33c}. & WinRAR CVE-2018-20250 \cite{apt33b}. For privilege escalation:  CVE-2017-0213 \cite{apt33a}. & Shamoon, TURNEDUP, AutoIt \cite{apt33b} and POWERTON \cite{apt33a}.  \\
      \\

    APT 34, a.k.a OilReg & Spearphishing attachment \cite{apt34c}.  & MS office: CVE-2017-11882, Malicious RTF CVE-2017-0199 \cite{apt34c}. & ISMAgent malware Plink utility to create tunnels to \cc\ server \cite{apt34d}   \\ 

\\

     APT 35, a.k.a Majic Hound & Spearphishing attachment and link to deliver Pupy RAT or MagicHound.Rollover \cite{apt35a}. & N/A & IRC bot (MagicHound.Leash) MPK.trojan.rat\cite{apt35a}.   \\
\\
    APT 37, a.k.a ScarCruft & Spearphishing attachment and link Torrent (KARAE) \cite{apt37}.  & Adobe Flash: CVE-2016-4117, CVE-2018-4878. MS Word: CVE-2017-019 \cite{apt37}. & NavRAT, HAPPYWORK, Final1stspy, DOGCALL, CORALDECK, ROKRAT, SHUTTERSPEED, SLOWDRIFT, WINERACK and KARAE  \cite{apt37}.   \\
\\
   admin@338 & Spearphishing attachment (LOWBALL malware) \cite{admin338}. &N/A & Bubblewarp, LOWBALL and	PoisonIvy and Custom (PoisonIvy) \cite{admin338}. 	\\\\

      Blockbuster a.k.a Operation Flame, Lazarus Group, HIDDEN COBRA & Spearphishing attachment \cite{LazarusGroupB}. & N/A & HOPLIGHT, KEYMARBLE, HARDRAIN, FALLCHILL.rat, Bankshot.rat, BADCALL, WannaCry, Volgmer and Proxysvc \cite{LazarusGroupC}.   \\
\\
     Cobalt Group  a.k.a Cobalt Spider & Spearphishing attachment \cite{CobaltGroupA}. & CVE-2017-11882, IE: CVE-2018-8174, CVE-2017-8570, CVE-2017-0199, CVE-2017-8759 \cite{CobaltGroupB}. MS Word:  CVE-2015-1641 , Adobe Flash:  CVE-2016-4117 \cite{CobaltGroupA}. & Odinaff, $Odinaff\!g1$, $Odinaff\!gm$, Batel, Gussdoor, Ammyy \cite{CobaltGroupA} and  More\_eggs \cite{CobaltGroupB}.\\
\\
   
		\bottomrule 
    \end{supertabular}
    }
  \end{center}
\end{landscape}

\begin{landscape}
  \begin{center}
    {\footnotesize
    \setlength\tabcolsep{1pt}

    \tablefirsthead{%
      \toprule
      \makecell{\multirow{1}{*}{\textbf{\makecell{APT Campaign}}}} 
        & \makecell{\multirow{1}{*}{\textbf{\makecell{Delivery Method}}}}  
        & \makecell{\multirow{1}{*}{\textbf{\makecell{Vulnerability}}}} 
        & \makecell{\multirow{1}{*}{\textbf{\makecell{Backdoors}}}} \\
      \midrule
      \midrule
    }

    \tablehead{%
      \multicolumn{4}{l}{\tablename\ \thetable\ -- (Continued)} \\
      \toprule
      \makecell{\multirow{1}{*}{\textbf{\makecell{APT Campaign}}}} 
        & \makecell{\multirow{1}{*}{\textbf{\makecell{Delivery Method}}}}  
        & \makecell{\multirow{1}{*}{\textbf{\makecell{Vulnerability}}}} 
        & \makecell{\multirow{1}{*}{\textbf{\makecell{Backdoors}}}} \\
      \midrule
      \midrule
    }

    \tabletail{%
      \bottomrule
      \multicolumn{4}{r}{\textit{Continued on the next page}} \\
    }

    \tablelasttail{%
      \bottomrule
    }


    \begin{supertabular}{L{3cm}*{1}{L{4cm}}*{1}{L{6cm}}*{1}{L{10cm}}}

     Dragonfly 2.0 & Spam campaign  Waterhole attack, i.e. iFrame is used forward to another website hosting Lightsout \cite{Dragonfly2c}. & With using Lightsout exploit kit against Java 6: CVE-2012-1723 Java 7: CVE-2013-2465 and IE: CVE-2012-4792 CVE-2013-1347 \cite{Dragonfly2c}. & Backdoor.Oldrea, Trojan.Karagany \cite{Dragonfly2c}.  \\
\\
      Duqu 2.0 & Spearphishing attachment \cite{aptDuquA}. & MS Word with embedded TTF: CVE-2011-3402 $\dagger$, CVE-2014-4148 $\dagger$. For lateral movement: CVE-2014-6324  $\dagger$ \cite{aptDuquA}. & N/A  \\	
\\
      Elderwood,   a.k.a Operation Aurora,  Sneaky Panda  & Waterhole (IFRAME) to forward to the server that hosting exploit \cite{aptElderwoodA}.   & CVE-2012-0779 $\dagger$,  CVE-2012-1875 $\dagger$,  CVE-2012-1889 $\dagger$, CVE-2012-1535 $\dagger$, CVE-2011-0609 $\dagger$, CVE-2011-0611 $\dagger$, CVE-2011-2110 $\dagger$,  CVE-2010-0249 $\dagger$ \cite{aptElderwoodA}. & Backdoor.Briba, Backdoor.Ritsol, Backdoor.Nerex, Backdoor.Linfo, Backdoor.Wiarp, Backdoor.Vasport, Backdoor.Darkmoon, Trojan.Hydraq (Aurora), Trojan.Naid, Trojan.Pasam, Packed.Generic.379, Packed.Generic.374 \cite{aptElderwoodA}. \\
  \\    
	  FIN 7, a.k.a Carbanak & Spearphishing attachment malicious DOCX and RTF \cite{aptFIN7}. & N/A & Carbanak, POWERSOURCE, TEXTMATE and HALFBAKED \cite{aptFIN7}.   \\
          \\ 
 Leviathan,  a.k.a APT 40 & Spearphishing attachment and link \cite{apt40a}. &N/A & AIRBREAK, FRESHAIR, BEACON, Gh0st, PHOTO (Derusbi) , BADFLICK , China Chopper, PluX (Sogu) \cite{apt40a},  Cobalt Strike and BLACKCOFFEE \cite{apt40c}.   \\
\\
      Naikon APT & Spearphishing attachment to drop naikon backdoor \cite{aptNaikonA}. & CVE-2012-0158, CVE-2010-3333 \cite{aptNaikonA}. & Naikon backdoor, RARSTONE, SslMM, Sys10, xsPlus, MsnMM, Sakto \cite{aptNaikonA}, WinMM, WininetMM \cite{aptNaikonB}.   \\
\\
      Patchwork a.k.a Dropping Elephant  MONSOON & Spearphishing attachment and link  Drive-by-Download i.e. Adobe Flash Update  \cite{aptPatchworkB}. & CVE-2014-4114, CVE-2012-1856, CVE-2017-0199, CVE-2017-8570, CVE-2015-1641 to deliver Badnews backdoor \cite{aptPatchworkB}. & AutoIt, Unknown Logger \cite{aptPatchworkB}, QuasarRAT \cite{aptPatchworkD}, 	TINYTYPHON \cite{aptPatchworkE}, Socksbot, NDiskMonitor and Badnews \cite{aptPatchworkB}.   \\
\\
      Sandworm Team a.k.a BlackEnergy Quedagh \cite{aptSandwormB}.& Spearphishing attachment \cite{aptSandwormB}. &  CVE-2010-3333 \cite{aptSandwormB}. & BlackEnergy \cite{aptSandwormB}.   \\
\\
      Strider a.k.a PROJECTSAURON \cite{aptStriderB}. & Plugged USB sticks \cite{aptStriderB}. & N/A & Remsec: Loader: MSAOSSPC.DLL Lua modules: Pip backdoor and HTTP backdoor  \cite{aptStriderA}.   \\
\\
     Regin & Spearphishing attachment \cite{aptReginA}. & N/A & Backdoor.Regin \cite{aptReginA}.  \\
\\
      Red October, a.k.a Cloud Atlas \cite{aptRedOctoberc}. & Spearphishing attachment \cite{aptRedOctobera}. & MS Excel: CVE-2009-3129, MS Word: CVE-2010-3333, CVE-2012-0158, Rhino Java: CVE-2011-3544 \cite{aptRedOctobera}. & LHAFD.GCP \cite{aptRedOctobera}, Zakladka a.k.a winupdate.dll and  WNFTPSCAN \cite{aptRedOctoberb}.  \\
   
		\bottomrule 
    \end{supertabular}
    }
  \end{center}
\end{landscape}

\begin{landscape}
  \begin{center}
    {\footnotesize
    \setlength\tabcolsep{1pt}

    \tablefirsthead{%
      \toprule
      \makecell{\multirow{6}{*}{\textbf{\makecell{APT \\ Campaign}}}} 
        & \multicolumn{2}{c}{\textbf{\makecell{Evasion Tech.}}} & 
        & \multicolumn{5}{c}{\textbf{Other \cc\ Settings}} 
        & \makecell{\multirow{6}{*}{\textbf{\makecell{Note}}}} \\
      \cmidrule(lr){2-3} \cmidrule(lr){5-9}
        & \makecell{\multirow{4}{*}{\makecell{DNS}}} 
        & \makecell{\multirow{4}{*}{\makecell{HTTP}}} 
        &
        & \makecell{\multirow{4}{*}{\makecell{Channels}}} 
        & \makecell{\multirow{4}{*}{\makecell{Protocols}}} 
        & \makecell{\multirow{4}{*}{\makecell{Obfuscation}}}
        & \makecell{\multirow{4}{*}{\makecell{Payload}}} 
        & \makecell{\# FQDNs \\ / \cc\ IPs}\\
      \midrule
      \midrule
    }

    \tablehead{%
      \multicolumn{10}{l}{\tablename\ \thetable\ -- (Continued)}\\
      \toprule
      \makecell{\multirow{6}{*}{\textbf{\makecell{APT \\ Campaign}}}} 
        & \multicolumn{2}{c}{\textbf{\makecell{Evasion Tech.}}} & 
        & \multicolumn{5}{c}{\textbf{Other \cc\ Settings}} 
        & \makecell{\multirow{6}{*}{\textbf{\makecell{Note}}}} \\
      \cmidrule(lr){2-3} \cmidrule(lr){5-9}
        & \makecell{\multirow{4}{*}{\makecell{DNS}}} 
        & \makecell{\multirow{4}{*}{\makecell{HTTP}}} 
        &
        & \makecell{\multirow{4}{*}{\makecell{Channels}}} 
        & \makecell{\multirow{4}{*}{\makecell{Protocols}}} 
        & \makecell{\multirow{4}{*}{\makecell{Obfuscation}}}
        & \makecell{\multirow{4}{*}{\makecell{Payload}}} 
        & \makecell{\# FQDNs \\ / \cc\ IPs}\\
      \midrule
      \midrule
    }

    \tabletail{%
      \bottomrule
      \multicolumn{10}{r}{\textit{Continued on the next page}}\\
    }

    \tablelasttail{%
      \bottomrule
    }

    \topcaption{APT Campaigns From Network Perspective - Part II.}
    \label{tab:long_table2}

    \begin{supertabular}{L{2cm}*{1}{L{1.5cm}}*{1}{L{3cm}}*{1}{L{0.2cm}}%
                         *{1}{L{5.0cm}}*{1}{L{1.5cm}}*{1}{L{3cm}}*{1}{L{3.0cm}}%
                         *{1}{L{1.0cm}}*{1}{L{3.0cm}}}

  APT 1, a.k.a Comment Crew  & Hijacked FQDNs, Dynamic DNS \cite{apt1}. & HTTPS, Mimicked MSN Messenger, Jabber/XMPP, Gmail Calendar \cite{apt1}. & &\cc\ Over HTTP, \cc\ over DNS \cite{apt1}. & HTTP, HTTPS, FTP \cite{apt1}. & Base64, single-byte XOR, TLS, SSL and 3DES \cite{apt1}. & GDOCUPLOAD, GETMAIL, LIGHTBOLT, MAPIGET \cite{apt1}.& (2551)/ (849) \cite{apt1}. & In addtion, APT 1 establish >937 \cc\ servers \cite{apt1}.\\
  \\
      
       APT 2, a.k.a Putter Panda  & Dynamic DNS \cite{apt2b}. & N/A & & abc.scr, a dropper, installs 4H RAT to open the first channel to connect to \cc\ server. The next channel is initiated with 3PARA followed by several channels established by Httpclient and Pngdowner \cite{apt2a}.  &HTTP \cite{apt2a}\cite{apt2b}.& RC4, 16-byte XOR, DES (CBC) With MD5 hash key of a string in HTTP request back to \cc\ 1-byte XOR  with 0xBE \cite{apt2a}. & N/A &(>57)/ (>32) \cite{apt2a}.&  \\
       \\
      APT 3, a.k.a Gothic Panda a.k.a Pirpi  Buckeye  & N/A & HTTP Proxy and HTTP Cookie field \cite{apt3e}. & & The main channels established by Pirpi \cite{apt3d}. & HTTP, HTTPS, SOCKS5 and FTP \cite{apt3d}.  & N/A & Keylogger,  RemoteCMD, PwDumpVariant, OSinfo, ChromePass, Lazagne \cite{apt3b}, Customized Mimikatz, Dsquery \cite{apt3e}. & (5)/ (2) \cite{apt3a}. &  \\
        \\
      APT 10 \\ a.k.a menuPass  & Dynamic DNS \cite{apt10a}. & Cookies embedding in HTTP, HTTPS \cite{apt10b}.  &&  Through legitimate access to many Managed IT Service Providers (MSP) \cite{apt10a}, or embedding data in cookies field in HTTP header \cite{apt10b}. & HTTP, HTTPS, FTP \cite{apt10b}. & AES, RC4, MD5 and Base64-encoding \cite{apt10b}. & MimiKatz, PwDump6 and certutil \cite{apt10b}. & (102)/ (25)\cite{apt10a}.&  \\
        \\
      APT 12  &N/A& N/A &&Over HTTP \cite{apt12}. &HTTP \cite{apt12}. & RC4 \cite{apt12}.&N/A&N/A& \\
      \\
      APT 15, a.k.a Ke3chang  Vixen Panda  & Dynamic DNS \cite{apt15a} and RoyalDNS uses DNS for \cc\ \cite{apt15b}. & HTTP with COM interface IWebBrowser2 \cite{apt15b}. && Third Party DNS Service "Nwsapagent"\cite{apt15b}. & RoyalDNS uses TXT Record in DNS protocol to send payloads \cite{apt15b}. & & Spwebmember, custom keylogger, Mimikatz, other network scanning and enumeration tools \cite{apt15b}. & (11)/ (22)\cite{apt15a}.\\
      \\
      
		\bottomrule 
    \end{supertabular}
    }
  \end{center}
\end{landscape}

\begin{landscape}
  \begin{center}
    {\footnotesize
    \setlength\tabcolsep{1pt}

    \tablefirsthead{%
      \toprule
      \makecell{\multirow{6}{*}{\textbf{\makecell{APT \\ Campaign}}}} 
        & \multicolumn{2}{c}{\textbf{\makecell{Evasion Tech.}}} & 
        & \multicolumn{5}{c}{\textbf{Other \cc\ Settings}} 
        & \makecell{\multirow{6}{*}{\textbf{\makecell{Note}}}} \\
      \cmidrule(lr){2-3} \cmidrule(lr){5-9}
        & \makecell{\multirow{4}{*}{\makecell{DNS}}} 
        & \makecell{\multirow{4}{*}{\makecell{HTTP}}} 
        &
        & \makecell{\multirow{4}{*}{\makecell{Channels}}} 
        & \makecell{\multirow{4}{*}{\makecell{Protocols}}} 
        & \makecell{\multirow{4}{*}{\makecell{Obfuscation}}}
        & \makecell{\multirow{4}{*}{\makecell{Payload}}} 
        & \makecell{\# FQDNs \\ / \cc\ IPs}\\
      \midrule
      \midrule
    }

    \tablehead{%
      \multicolumn{10}{l}{\tablename\ \thetable\ -- (Continued)}\\
      \toprule
      \makecell{\multirow{6}{*}{\textbf{\makecell{APT \\ Campaign}}}} 
        & \multicolumn{2}{c}{\textbf{\makecell{Evasion Tech.}}} & 
        & \multicolumn{5}{c}{\textbf{Other \cc\ Settings}} 
        & \makecell{\multirow{6}{*}{\textbf{\makecell{Note}}}} \\
      \cmidrule(lr){2-3} \cmidrule(lr){5-9}
        & \makecell{\multirow{4}{*}{\makecell{DNS}}} 
        & \makecell{\multirow{4}{*}{\makecell{HTTP}}} 
        &
        & \makecell{\multirow{4}{*}{\makecell{Channels}}} 
        & \makecell{\multirow{4}{*}{\makecell{Protocols}}} 
        & \makecell{\multirow{4}{*}{\makecell{Obfuscation}}}
        & \makecell{\multirow{4}{*}{\makecell{Payload}}} 
        & \makecell{\# FQDNs \\ / \cc\ IPs}\\
      \midrule
      \midrule
    }

    \tabletail{%
      \bottomrule
      \multicolumn{10}{r}{\textit{Continued on the next page}}\\
    }

    \tablelasttail{%
      \bottomrule
    }

    \begin{supertabular}{L{2cm}*{1}{L{1.5cm}}*{1}{L{3cm}}*{1}{L{0.2cm}}%
                         *{1}{L{5.0cm}}*{1}{L{1.5cm}}*{1}{L{3cm}}*{1}{L{3.0cm}}%
                         *{1}{L{1.0cm}}*{1}{L{3.0cm}}}

      APT 16, a.k.a EPS  &N/A& HTTP Beacon and HTTPS \cite{apt16a}. && Variants of ELMER  communicate with two different \cc\ locations \cite{apt16a}.& HTTP and ELMER beacons over HTTPS \cite{apt16a}.& N/A &  N/A& (2)/ (2) \cite{apt16a}. &  \\
      \\
      APT 17 a.k.a Deputy Dog  & N/A& N/A && \cc\ distribution over Microsoft TechNet and Social Media\cite{apt17}. &HTTP \cite{apt17}.&N/A&N/A&N/A& BLACKCOFFEE holds a URL for the actor profile or thread at Microsoft TechNet or Social media to retrieve \cc\ IP address which will be updated once it is exposed\cite{apt17}. \\		
\\
      APT 18, a.k.a Wekby, Dynamite Panda & DNS as a covert channel \cite{apt18a}. & HTTPS \cite{aptGhost}.  && HcdLoader used for lateral movement and data Exfiltration over HTTP, HTTPBrowser, and Pisloader used DNS requests as a channel in TXT records \cite{apt18a}. & HTTP, HTTPS \cite{aptGhost} and DNS \cite{apt18d}. & Pisloader uses base32-encoded \cite{apt18d}. &HTTPBrowser sends Keystrokes \cite{apt18b}. & N/A& Xyligan establish the foothold, then install Hcdloader to provide command-line access \cite{apt18b}. \\
      \\
      APT 19 a.k.a. Codoso & N/A&N/A  &&HTTP over port 22, SCP, SFTP over port 22, HTTPS and DNS \cite{apt19b}& HTTP  \cite{apt19b} & Base64, single-byte XOR keys \cite{apt19b}. &Cobalt Strike \cite{apt19b}.& (4)/ (4) \cite{apt19}.&  \\
\\
         APT 27 a.k.a Emissary Panda, Threat Group-3390, LuckyMouse &N/A &N/A  && N/A & HTTP, HTTPS and DNS \cite{apt27a}.& Base64 Encoding, Metasploit’s Shikata Ganai, encoder and LZNT1 \cite{apt27b}. & ASPXSpy (webshell) 	China Chopper, OwaAuth (to steal Exchange's Passwords) and Mimikatz \cite{apt27c}. &(5) \cite{apt27b}/ (2) \cite{apt27c}. & Main \cc\ bbs.sonypsps[.]com with resolved IP, i.e. belong to Router, at Ukrainian ISP network, that was hacked to pass malware's HTTP request \cite{apt27b}.   \\\\

      APT 28, a.k.a Sednit Sofacy Fancy Bear & DGA used by Sofacy.WinHttp \cite{apt28c} & Custom (CHOPSTICK) over 80/443  \cite{apt28a}. && First Channel: Downdelph over HTTP \cite{apt28c}. 2nd Channel: CHOPSTICK over HTTP. 3rd Channel: CORESHELL \cite{apt28a}. and Zebrocy over HTTP, SMTP, and POP3. Komplex and JHUHUGIT use HTTP Post and HTTPS while XTunnel uses SSL/TLS with RC4 \cite{apt28f}.& HTTP, HTTPS, SMTP and POP3 \cite{apt28a,apt28e,apt28f}. & CORESHELL uses Base64 encoding \cite{apt28a}. & N/A & N/A& APT 28 uses POP3 to communicate with GMAIL services to allocate FQDNs and \cc\ locations \cite{apt28a}.\\

		\bottomrule 
    \end{supertabular}
    }
  \end{center}
\end{landscape}

\begin{landscape}
  \begin{center}
    {\footnotesize
    \setlength\tabcolsep{1pt}

    \tablefirsthead{%
      \toprule
      \makecell{\multirow{6}{*}{\textbf{\makecell{APT \\ Campaign}}}} 
        & \multicolumn{2}{c}{\textbf{\makecell{Evasion Tech.}}} & 
        & \multicolumn{5}{c}{\textbf{Other \cc\ Settings}} 
        & \makecell{\multirow{6}{*}{\textbf{\makecell{Note}}}} \\
      \cmidrule(lr){2-3} \cmidrule(lr){5-9}
        & \makecell{\multirow{4}{*}{\makecell{DNS}}} 
        & \makecell{\multirow{4}{*}{\makecell{HTTP}}} 
        &
        & \makecell{\multirow{4}{*}{\makecell{Channels}}} 
        & \makecell{\multirow{4}{*}{\makecell{Protocols}}} 
        & \makecell{\multirow{4}{*}{\makecell{Obfuscation}}}
        & \makecell{\multirow{4}{*}{\makecell{Payload}}} 
        & \makecell{\# FQDNs \\ / \cc\ IPs}\\
      \midrule
      \midrule
    }

    \tablehead{%
      \multicolumn{10}{l}{\tablename\ \thetable\ -- (Continued)}\\
      \toprule
      \makecell{\multirow{6}{*}{\textbf{\makecell{APT \\ Campaign}}}} 
        & \multicolumn{2}{c}{\textbf{\makecell{Evasion Tech.}}} & 
        & \multicolumn{5}{c}{\textbf{Other \cc\ Settings}} 
        & \makecell{\multirow{6}{*}{\textbf{\makecell{Note}}}} \\
      \cmidrule(lr){2-3} \cmidrule(lr){5-9}
        & \makecell{\multirow{4}{*}{\makecell{DNS}}} 
        & \makecell{\multirow{4}{*}{\makecell{HTTP}}} 
        &
        & \makecell{\multirow{4}{*}{\makecell{Channels}}} 
        & \makecell{\multirow{4}{*}{\makecell{Protocols}}} 
        & \makecell{\multirow{4}{*}{\makecell{Obfuscation}}}
        & \makecell{\multirow{4}{*}{\makecell{Payload}}} 
        & \makecell{\# FQDNs \\ / \cc\ IPs}\\
      \midrule
      \midrule
    }

    \tabletail{%
      \bottomrule
      \multicolumn{10}{r}{\textit{Continued on the next page}}\\
    }

    \tablelasttail{%
      \bottomrule
    }


    \begin{supertabular}{L{2cm}*{1}{L{1.5cm}}*{1}{L{3cm}}*{1}{L{0.2cm}}%
                         *{1}{L{5.0cm}}*{1}{L{1.5cm}}*{1}{L{3cm}}*{1}{L{3.0cm}}%
                         *{1}{L{1.0cm}}*{1}{L{3.0cm}}}

      APT 29, a.k.a CozyBear CozyDuke &N/A & Domain Fronting over HTTPS places malicious domains in the HTTP header and legitimate ones in the TLS header  \cite{apt29d}. && First CozyCar over HTTP/HTTPS, Second PowerShell Script, Third (SeaDaddy) over 443 \cite{apt29a}, Tor-meek for domain fronting and Multihop Proxy \cite{apt29d}. & HTTP, HTTPS, RDP (3389), NetBios (139), SMB (445), FTP  \cite{apt29b}. & Base64 Encoding \cite{apt29c}.& Backdoors for importing persistent PowerShell scripts \cite{apt29d} and exfiltrating data \cite{apt29c}. & (18)/(31) \cite{apt29b}. & CozyCar is embedded with an alternative \cc\ channel to Twitter, MiniDuke uses Google Search if Twitter approach \cite{apt29b} failed.  \\\\
    
     APT 30 & N/A & NETEAGLE over HTTP proxy post request, or UDP (6000) BACKSPACE disable firewall \cite{apt30}. && NETEAGLE connect with proxy over HTTP post beacons. If failed, RDP over TCP (7519) \cite{apt30}. BACKSPACE uses one domain to receive an update and another one for a backup with two more run-hide configurations \cite{apt30}. & HTTP and  RDP \cite{apt30}. & FLASH- FLOOD uses zlib, byte-rotation, and XOR NETEAGLE uses RC4 \cite{apt30}. & SPACESHIP and SHIPSHAPE exfiltrate data \cite{apt30}. & (5>) / (N/A)  \cite{apt30}. & SHIPSHAPE provides persistence by propagating through removable devices of the infected network against air-gap setting \cite{apt30}. \\
\\

      APT 32, a.k.a OceanLotus  & DNS tunneling for \cc\ and Data Exfiltration \cite{apt32a, apt32c}.  & Custom (Outlook macro backdoor over SMTP \cite{apt32a}), (JavaScript over HTTP 14146 \cite{apt32b} or HTTPS \cite{apt32f}. && Denis \cite{apt32a} and SOUNDBITE exfiltrate over DNS packets while PHOREAL uses ICMP and WINDSHIELD  communicate over raw TCP sockets \cite{apt32e}.  & HTTP, HTTPS \cite{apt32f} , DNS \cite{apt32a} , P2P over SMB and ICMP \cite{apt32e}. & AES-256(CBC) \cite{apt32b},  RC4 \cite{apt32d},  RSA256 and Base64 encoding \cite{apt32e}. & Mimikatz and data exfiltration \cite{apt32a}. &(62)/(22) \cite{apt32d}.& Backdoors exploits Microsoft Outlook \cite{apt32a}.	   \\
      \\
      APT 33, a.k.a Elfin  Shamoon  & N/A & HTTP Proxy \cite{apt33a}. && Shamoon uses HTTP, PoshC2 may uses proxy to \cc\ server over HTTP/HTTPS \cite{apt33a} and NanoCore over 6666 \cite{apt33c} & HTTP, HTTPS and FTP \cite{apt33b}. & AES, base64 encoding, DES \cite{apt33a}. & SniffPass, DarkComet, ProcDump, Mimikatz, PoshC2 and data exfiltration \cite{apt33b}. & (69) / (69) \cite{apt33b}. & Shamoon is responsible for  C\&C as well as data destruction processes. PoshC2 has multiple functions, including C\&C, search for passwords, Netstat, and pass the hash to gain access without plaintext \cite{apt33b}. \\
      \\

		\bottomrule 
    \end{supertabular}
    }
  \end{center}
\end{landscape}

\begin{landscape}
  \begin{center}
    {\footnotesize
    \setlength\tabcolsep{1pt}

    \tablefirsthead{%
      \toprule
      \makecell{\multirow{6}{*}{\textbf{\makecell{APT \\ Campaign}}}} 
        & \multicolumn{2}{c}{\textbf{\makecell{Evasion Tech.}}} & 
        & \multicolumn{5}{c}{\textbf{Other \cc\ Settings}} 
        & \makecell{\multirow{6}{*}{\textbf{\makecell{Note}}}} \\
      \cmidrule(lr){2-3} \cmidrule(lr){5-9}
        & \makecell{\multirow{4}{*}{\makecell{DNS}}} 
        & \makecell{\multirow{4}{*}{\makecell{HTTP}}} 
        &
        & \makecell{\multirow{4}{*}{\makecell{Channels}}} 
        & \makecell{\multirow{4}{*}{\makecell{Protocols}}} 
        & \makecell{\multirow{4}{*}{\makecell{Obfuscation}}}
        & \makecell{\multirow{4}{*}{\makecell{Payload}}} 
        & \makecell{\# FQDNs \\ / \cc\ IPs}\\
      \midrule
      \midrule
    }

    \tablehead{%
      \multicolumn{10}{l}{\tablename\ \thetable\ -- (Continued)}\\
      \toprule
      \makecell{\multirow{6}{*}{\textbf{\makecell{APT \\ Campaign}}}} 
        & \multicolumn{2}{c}{\textbf{\makecell{Evasion Tech.}}} & 
        & \multicolumn{5}{c}{\textbf{Other \cc\ Settings}} 
        & \makecell{\multirow{6}{*}{\textbf{\makecell{Note}}}} \\
      \cmidrule(lr){2-3} \cmidrule(lr){5-9}
        & \makecell{\multirow{4}{*}{\makecell{DNS}}} 
        & \makecell{\multirow{4}{*}{\makecell{HTTP}}} 
        &
        & \makecell{\multirow{4}{*}{\makecell{Channels}}} 
        & \makecell{\multirow{4}{*}{\makecell{Protocols}}} 
        & \makecell{\multirow{4}{*}{\makecell{Obfuscation}}}
        & \makecell{\multirow{4}{*}{\makecell{Payload}}} 
        & \makecell{\# FQDNs \\ / \cc\ IPs}\\
      \midrule
      \midrule
    }

    \tabletail{%
      \bottomrule
      \multicolumn{10}{r}{\textit{Continued on the next page}}\\
    }

    \tablelasttail{%
      \bottomrule
    }


    \begin{supertabular}{L{2cm}*{1}{L{1.5cm}}*{1}{L{3cm}}*{1}{L{0.2cm}}%
                         *{1}{L{5.0cm}}*{1}{L{1.5cm}}*{1}{L{3cm}}*{1}{L{3.0cm}}%
                         *{1}{L{1.0cm}}*{1}{L{3.0cm}}}

    APT 34, a.k.a OilReg & DNS tunneling \cite{apt34d} and DGA \cite{apt34c}. & N/A &&  First 'dolt' function in the PowerShell script initiates the first channel, Then ISMAgent uses DNS tunneling instead of HTTP \cite{apt34d}.  & HTTP, HTTPS, DNS, FTP \cite{apt34b}. & Cryptographic Data Encoding \cite{apt34b}.  & keylogger.KEYPUNCH, screenshot.CANDYKING, Tool.Plink to create tunnels Tool.netscan.SoftPerfect, Tool.netscan.GOLDIRONY\cite{apt34b}. & (10) / (14)  \cite{apt34b, apt34a}. &   \\ 

\\

     APT 35, a.k.a Majic Hound  & N/A & HTTPS \cite{apt35a}. && Over 4443, 3543 \cite{apt35a}. & HTTP, HTTPS, FTP, IRC, and SOAP \cite{apt35b}. & base64, AES\cite{apt35a}.& Pupy variant of Mimikats, CWoolger for keylogging, FireMalv for stealing credentials of Firefox Data Exfiltration \cite{apt35a}.& (39)/(97) \cite{apt35a}.&   \\
\\
    APT 37, a.k.a ScarCruft & N/A  & HTTP POST headers for data exfiltration \cite{apt37}. && First, HAPPYWORK and KARAE connect with \cc\ to receive further backdoors and open further channels CORALDECK connects with \cc\ through HTTP POST headers DOGCALL communicates with \cc\ and exfiltrates through cloud services, NavRAT uses SMTP POORAIM overall AOL Instant Messenger for \cc\ ROKRAT uses a variety of \cc\ channels including HTTPS and cloud services. SLOWDRIFT is used for cloud services \cite{apt37}. & HTTP, HTTPS, SMTP and P2P \cite{apt37}. & CORALDECK uses RAR protected with password DOGCALL uses single-byte XOR Finalspy uses Base64 Encoding \cite{apt37}. &N/A &N/A & Not only APT 37 exfiltrate credential data but also the microphone data, snapshot of virtual machines \cite{apt37}.   \\
\\
   admin@338  &N/A & N/A&& First, LOWBALL open a \cc\ channel over HTTPS  and can also use Cloud services \cite{admin338}. The next level for \cc\ traffic is performed by BUBBLEWRAP over HTTP/HTTPS or SOCKS \cite{admin338}. &N/A&N/A& HTTP, HTTPS and SOCKS \cite{admin338}. &N/A& 	\\\\

      Blockbuster, a.k.a Operation Flame, Lazarus Group, HIDDEN COBRA & N/A & HTTP  HTTPS (SSL) HTTPS (fake TLS) BADCALL disables Windows firewall \cite{LazarusGroupC}. && WannaCry uses Tor for \cc\ communication Volgmer, TYPEFRAME, Proxysvc, KEYMARBLE, BADCALL uses HTTPS. However, TYPEFRAME, AuditCred, and BADCALL connect to a proxy server RATANKBA and Proxysvc are used for data exfiltration  KEYMARBLE has multiple channels for exfiltration  Bankshot uses HTTP channel for exfiltration \cite{LazarusGroupC}. & HTTP, SMTP and RDP \cite{LazarusGroupC}. & Symmetric stream cipher e.g. AES, RC4 and Caracachs and XOR \cite{LazarusGroupA}. & The main payload is for data exfiltration. In addition, customized password dump tools, and batch scripts are uploaded \cite{LazarusGroupB}. & (2) /(3) \cite{LazarusGroupE}. & HARDRAIN, FALLCHILL, BADCALL use fake TLS Bankshot to create a fake TLS handshaking with the use of a public certificate \cite{LazarusGroupC}. 45 unique malware families during this operation \cite{LazarusGroupB}.  \\\\

		\bottomrule 
    \end{supertabular}
    }

  \end{center}
\end{landscape}

\begin{landscape}
  \begin{center}
    {\footnotesize
    \setlength\tabcolsep{1pt}

    \tablefirsthead{%
      \toprule
      \makecell{\multirow{6}{*}{\textbf{\makecell{APT \\ Campaign}}}} 
        & \multicolumn{2}{c}{\textbf{\makecell{Evasion Tech.}}} & 
        & \multicolumn{5}{c}{\textbf{Other \cc\ Settings}} 
        & \makecell{\multirow{6}{*}{\textbf{\makecell{Note}}}} \\
      \cmidrule(lr){2-3} \cmidrule(lr){5-9}
        & \makecell{\multirow{4}{*}{\makecell{DNS}}} 
        & \makecell{\multirow{4}{*}{\makecell{HTTP}}} 
        &
        & \makecell{\multirow{4}{*}{\makecell{Channels}}} 
        & \makecell{\multirow{4}{*}{\makecell{Protocols}}} 
        & \makecell{\multirow{4}{*}{\makecell{Obfuscation}}}
        & \makecell{\multirow{4}{*}{\makecell{Payload}}} 
        & \makecell{\# FQDNs \\ / \cc\ IPs}\\
      \midrule
      \midrule
    }

    \tablehead{%
      \multicolumn{10}{l}{\tablename\ \thetable\ -- (Continued)}\\
      \toprule
      \makecell{\multirow{6}{*}{\textbf{\makecell{APT \\ Campaign}}}} 
        & \multicolumn{2}{c}{\textbf{\makecell{Evasion Tech.}}} & 
        & \multicolumn{5}{c}{\textbf{Other \cc\ Settings}} 
        & \makecell{\multirow{6}{*}{\textbf{\makecell{Note}}}} \\
      \cmidrule(lr){2-3} \cmidrule(lr){5-9}
        & \makecell{\multirow{4}{*}{\makecell{DNS}}} 
        & \makecell{\multirow{4}{*}{\makecell{HTTP}}} 
        &
        & \makecell{\multirow{4}{*}{\makecell{Channels}}} 
        & \makecell{\multirow{4}{*}{\makecell{Protocols}}} 
        & \makecell{\multirow{4}{*}{\makecell{Obfuscation}}}
        & \makecell{\multirow{4}{*}{\makecell{Payload}}} 
        & \makecell{\# FQDNs \\ / \cc\ IPs}\\
      \midrule
      \midrule
    }

    \tabletail{%
      \bottomrule
      \multicolumn{10}{r}{\textit{Continued on the next page}}\\
    }

    \tablelasttail{%
      \bottomrule
    }


    \begin{supertabular}{L{2cm}*{1}{L{1.5cm}}*{1}{L{3cm}}*{1}{L{0.2cm}}%
                         *{1}{L{5.0cm}}*{1}{L{1.5cm}}*{1}{L{3cm}}*{1}{L{3.0cm}}%
                         *{1}{L{1.0cm}}*{1}{L{3.0cm}}}

     Cobalt Group a.k.a Cobalt Spider & DNS tunneling \cite{CobaltGroupA}. & HTTPS \cite{CobaltGroupA}. && Cobalt Groups uses Plink to open an SSH channel. In addition, the group uses Cobalt Strike for a variety of channels, i.e., HTTP, HTTPS, DNS, and VNC \cite{CobaltGroupC} over remote framebuffer (RBF Protocol), and it can send over one channel and received from another channel & HTTPS, DNS and P2P SMB \cite{CobaltGroupA}.  &  N/A & Cobalt Strike is responsible for multiple payloads, including keylogging, PowerShell scripts, and data exfiltration \cite{CobaltGroupA}. &N/A& \\
\\
     Dragonfly 2.0  & N/A & Backdoor.Oldrea uses base64 encoded string of HTTP Get or RSA with HTTP Post \cite{Dragonfly2c}. && First, Backdoor.Oldrea communicates with \cc\ server. Second, Trojan.Karagany uses a live connection to Microsoft or Adobe websites if available \cite{Dragonfly2c}.  &HTTP \cite{Dragonfly2c}.&RSA, Base64 Encoding and XOR \cite{Dragonfly2c}.& Trojan.Karagany is implanted to conduct internal reconnaissance, then Backdoor.Oldrea exfiltrates Credentials, Emails, OWA address book, processes, and infrastructure info. and classified documents \cite{Dragonfly2c} \cite{Dragonfly2b}. &(N/A) / (13) \cite{Dragonfly2a} & The group extends their work after Dragonfly 1.0 \cite{Dragonfly2c}. \\
\\
      Duqu 2.0 & N/A & Bidirectional HTTP Proxy, embed \cc\ traffic inside JPEG or GIF over HTTP or inside driver files of SMB with knocking mechanism for tunneling or fake TCP/IP packets to specific IP \cite{aptDuquA}. && Duqu 2.0 uses a variety of channels, including HTTP, HTTPS, and tunneling SMB/RDB network pipes over HTTPS\cite{aptDuquA}.  & Outside LAN: HTTPS  while inside LAN: SMB network pipes or RDP\cite{aptDuquA}. & Symmetric stream cipher e.g. AES\cite{aptDuquA} and symmetric block cipher e.g. Camellia 256 and XXTEA \cite{aptDuquA}. &Exfiltrate highly classified documents related to nuclear program\cite{aptDuquA}. & N/A & The threat actor penetrates a certificate authority in Hungary and is able to generate legitimate certificates\cite{aptDuquA}.  \\	
\\
      Elderwood,   a.k.a Operation Aurora,  Sneaky Panda   &N/A &N/A && All backdoors connect to a shared \cc\ infrastructure  \cite{aptElderwoodA}. &N/A&N/A&N/A&(18) / (N/A) \cite{aptElderwoodA}.  \\
  \\    
	  FIN 7, a.k.a Carbanak  & Embedding data in TXT record \cite{aptFIN7}. &N/A &&POWERSOURCE embed \cc\ traffic in TXT record of DNS packet \cite{aptFIN7}.  If POWERSOURCE is detected, then TEXTMATE opens another channel using the same technique, another \cc\ channel over GoogleDoc \cite{aptFIN7}. Finally, the Ammyy Admin tool, a legitimate tool, is used as \cc\ channel \cite{aptFIN7b}.  & Remote Desktop Protocol (RDP)HTTP  HTTPS  DNS \cite{aptFIN7}& N/A &N/A & N/A&  \\
          \\

		\bottomrule 
    \end{supertabular}
    }
  \end{center}
\end{landscape}

\begin{landscape}
  \begin{center}
    {\footnotesize
    \setlength\tabcolsep{1pt}

    \tablefirsthead{%
      \toprule
      \makecell{\multirow{6}{*}{\textbf{\makecell{APT \\ Campaign}}}} 
        & \multicolumn{2}{c}{\textbf{\makecell{Evasion Tech.}}} & 
        & \multicolumn{5}{c}{\textbf{Other \cc\ Settings}} 
        & \makecell{\multirow{6}{*}{\textbf{\makecell{Note}}}} \\
      \cmidrule(lr){2-3} \cmidrule(lr){5-9}
        & \makecell{\multirow{4}{*}{\makecell{DNS}}} 
        & \makecell{\multirow{4}{*}{\makecell{HTTP}}} 
        &
        & \makecell{\multirow{4}{*}{\makecell{Channels}}} 
        & \makecell{\multirow{4}{*}{\makecell{Protocols}}} 
        & \makecell{\multirow{4}{*}{\makecell{Obfuscation}}}
        & \makecell{\multirow{4}{*}{\makecell{Payload}}} 
        & \makecell{\# FQDNs \\ / \cc\ IPs}\\
      \midrule
      \midrule
    }

    \tablehead{%
      \multicolumn{10}{l}{\tablename\ \thetable\ -- (Continued)}\\
      \toprule
      \makecell{\multirow{6}{*}{\textbf{\makecell{APT \\ Campaign}}}} 
        & \multicolumn{2}{c}{\textbf{\makecell{Evasion Tech.}}} & 
        & \multicolumn{5}{c}{\textbf{Other \cc\ Settings}} 
        & \makecell{\multirow{6}{*}{\textbf{\makecell{Note}}}} \\
      \cmidrule(lr){2-3} \cmidrule(lr){5-9}
        & \makecell{\multirow{4}{*}{\makecell{DNS}}} 
        & \makecell{\multirow{4}{*}{\makecell{HTTP}}} 
        &
        & \makecell{\multirow{4}{*}{\makecell{Channels}}} 
        & \makecell{\multirow{4}{*}{\makecell{Protocols}}} 
        & \makecell{\multirow{4}{*}{\makecell{Obfuscation}}}
        & \makecell{\multirow{4}{*}{\makecell{Payload}}} 
        & \makecell{\# FQDNs \\ / \cc\ IPs}\\
      \midrule
      \midrule
    }

    \tabletail{%
      \bottomrule
      \multicolumn{10}{r}{\textit{Continued on the next page}}\\
    }

    \tablelasttail{%
      \bottomrule
    }


    \begin{supertabular}{L{2cm}*{1}{L{1.5cm}}*{1}{L{3cm}}*{1}{L{0.2cm}}%
                         *{1}{L{5.0cm}}*{1}{L{1.5cm}}*{1}{L{3cm}}*{1}{L{3.0cm}}%
                         *{1}{L{1.0cm}}*{1}{L{3.0cm}}}

 Leviathan,  a.k.a APT 40  & NanHaiShu uses Dynamic DNS With Third Party \cite{apt40b}. & HTTPS \cite{apt40a}. && Derusbi opens two channels, an HTTP beacon, and HTTPS channel over 31800, several channels established by Gh0st, AIRBREAK and FRESHAIR  \cite{apt40a}.  & HTTP, HTTPS \cite{apt40a}. & XOR&N/A&N/A& APT 40 develop several custom tools and malware, including HOMEFRY. China Chopper is implanted as a web shell to brute force passwords, uploads, and downloads files packed with UPX, and commands are sent over HTTP Post \cite{apt40a}. APT 40 targets VPN credentials \cite{apt40a}.  \\
\\
      Naikon APT  & Dynamic DNS \cite{aptNaikonB}. & HTTPS \cite{aptNaikonA}. && Naikon backdoor drops all other backdoors to deploy a variety of channels, RARSTONE, which opens the first channel over HTTPS, then SslMM is installed to send the keystrokes and footprinting info. over two channels, Sys10 open another channel over HTTP to collect local IP addresses WinMM opens two channels to add more persistence over HTTP \cite{aptNaikonA}. & HTTP, HTTPS, FTP and TFTP \cite{aptNaikonA}. &N/A &HDoor is uploaded to the victim for internal reconnaissance over SOCKS5 proxy service  FTP is used for data exfiltration \cite{aptNaikonA}. &N/A&  \\
\\
      Patchwork a.k.a Dropping Elephant  MONSOON   &  N/A  &   HTTPS  && QuasarRAT opens multiple channels over SOCKS5 proxy and FTP \cite{aptPatchworkD}, Socksbot opens another channel over SOCKS5 proxy to run Powershell scripts, BADNEWS uses RSS feeds and Github for \cc\ \cite{aptPatchworkB}, TINYTYPHON is mainly used for data exfiltration \cite{aptPatchworkE}.  & HTTP, HTTPS, RDP, FTP and SOCKS5 \cite{aptPatchworkA} & QuasarRAT, NDiskMonitor use AES AutoIt base64 encoding, TINYTYPHON \cite{aptPatchworkE} and BADNEWS uses XOR \cite{aptPatchworkB}. & NDiskMonitor collects usernames, files, and directories and sends them back to \cc\ server Unknown Logger spread through USB to disable security tools, records keystrokes, and collect usernames and IP addresses \cite{aptPatchworkB}. &N/A &  \\
\\
   
		\bottomrule 
    \end{supertabular}
    }
  \end{center}
\end{landscape}

\begin{landscape}
  \begin{center}
    {\footnotesize
    \setlength\tabcolsep{1pt}

    \tablefirsthead{%
      \toprule
      \makecell{\multirow{6}{*}{\textbf{\makecell{APT \\ Campaign}}}} 
        & \multicolumn{2}{c}{\textbf{\makecell{Evasion Tech.}}} & 
        & \multicolumn{5}{c}{\textbf{Other \cc\ Settings}} 
        & \makecell{\multirow{6}{*}{\textbf{\makecell{Note}}}} \\
      \cmidrule(lr){2-3} \cmidrule(lr){5-9}
        & \makecell{\multirow{4}{*}{\makecell{DNS}}} 
        & \makecell{\multirow{4}{*}{\makecell{HTTP}}} 
        &
        & \makecell{\multirow{4}{*}{\makecell{Channels}}} 
        & \makecell{\multirow{4}{*}{\makecell{Protocols}}} 
        & \makecell{\multirow{4}{*}{\makecell{Obfuscation}}}
        & \makecell{\multirow{4}{*}{\makecell{Payload}}} 
        & \makecell{\# FQDNs \\ / \cc\ IPs}\\
      \midrule
      \midrule
    }

    \tablehead{%
      \multicolumn{10}{l}{\tablename\ \thetable\ -- (Continued)}\\
      \toprule
      \makecell{\multirow{6}{*}{\textbf{\makecell{APT \\ Campaign}}}} 
        & \multicolumn{2}{c}{\textbf{\makecell{Evasion Tech.}}} & 
        & \multicolumn{5}{c}{\textbf{Other \cc\ Settings}} 
        & \makecell{\multirow{6}{*}{\textbf{\makecell{Note}}}} \\
      \cmidrule(lr){2-3} \cmidrule(lr){5-9}
        & \makecell{\multirow{4}{*}{\makecell{DNS}}} 
        & \makecell{\multirow{4}{*}{\makecell{HTTP}}} 
        &
        & \makecell{\multirow{4}{*}{\makecell{Channels}}} 
        & \makecell{\multirow{4}{*}{\makecell{Protocols}}} 
        & \makecell{\multirow{4}{*}{\makecell{Obfuscation}}}
        & \makecell{\multirow{4}{*}{\makecell{Payload}}} 
        & \makecell{\# FQDNs \\ / \cc\ IPs}\\
      \midrule
      \midrule
    }

    \tabletail{%
      \bottomrule
      \multicolumn{10}{r}{\textit{Continued on the next page}}\\
    }

    \tablelasttail{%
      \bottomrule
    }


    \begin{supertabular}{L{2cm}*{1}{L{1.5cm}}*{1}{L{3cm}}*{1}{L{0.2cm}}%
                         *{1}{L{5.0cm}}*{1}{L{1.5cm}}*{1}{L{3cm}}*{1}{L{3.0cm}}%
                         *{1}{L{1.0cm}}*{1}{L{3.0cm}}}
      Sandworm Team a.k.a BlackEnergy Quedagh \cite{aptSandwormB}. & N/A & HTTP, POST requests \cite{aptSandwormB}. && BlackEnergy is known to operate over multiple channels with \cite{aptSandwormB}.  &  HTTP POST requests \texttt{getp} and \texttt{plv} fields for plugin getpd and for binaries download \cite{aptSandwormB}. & Base64 Encoding \cite{aptSandwormB}. & The payload includes highly classified documents and layouts of Ukrainian SCADA and government. In addition, BlackEnergy captures keystrokes and obtains credentials\cite{aptSandwormB}. &N/A  & BlackEnergy has the ability to create botnets for destructive attacks DDoS \cite{aptSandwormB}.  \\
\\
      Strider a.k.a PROJECTSAURON \cite{aptStriderB}. & DNS tunneling & HTTP over ICMP && Remsec open four channels to maximize the persistence over DNS, HTTP, HTTPS, and SMTP with TCP, UDP, or ICMP \cite{aptStriderA}.  & HTTP, ICMP, DNS and SMTP \cite{aptStriderA}. & Remsec uses multiple encryption schemes including RC5(CBC), Remsec.Null session pipes AES(CBC) and RSA \cite{aptStriderB}. & Remsec obtains keystrokes with its module Sauron, network infrastructure layout performs ARP scanning, and exfiltrates data\cite{aptStriderA}. & HTTP backdoor holds several URLs to locate \cc\ servers \cite{aptStriderA}.&  \\
\\
     Regin  & N/A & HTTP over TCP, UDP, and ICMP \cite{aptReginA}, and HTTPS over proxy of other victims in internal networks \cite{aptReginB}. && Regin opens multiple channels among internal networks (P2P) over network pipes SMB and ICMP raw socket that communicate with the machines on the border, which are acting as a router to forward traffic to \cc\ servers over HTTP or HTTPS \cite{aptReginB}.  & HTTP, HTTPS, SMTP, SMB \cite{aptReginB} and ICMP raw socket \cite{aptReginA}. & XOR and RC5 \cite{aptReginA}. & Data exfiltration, Regin also obtains keystrokes and footprinting info \cite{aptReginB}. & (N/A)/ (4)\cite{aptReginB}. & \\
\\
      Red October, a.k.a Cloud Atlas \cite{aptRedOctoberc}.  & N/A  &  Zakladka uses POST request \cite{aptRedOctoberb} && Red October backdoors allow a victim to connect to \cc\ through a chain of proxies with different locations \cite{aptRedOctobera}. Recently, it can also open HTTPS channels to connect cloud services for \cc\ communications \cite{aptRedOctoberc}. WNFTPSCAN Exfiltrate data to remote FTP servers \cite{aptRedOctoberc}.  & HTTP, HTTPS and FTP \cite{aptRedOctoberb}. & Zakladka uses RC4 and base64 encoding & Data Exfiltration modules such as WNFTPSCAN, GetFileReg, and FileInfo do not interact with \cc\ server directly  \cite{aptRedOctoberb}. &  (>60) / (10) \cite{aptRedOctobera}.& \\
   
		\bottomrule 
    \end{supertabular}
    }
  \end{center}
\end{landscape}

\begin{landscape}

\begin{table}[h!]
  \begin{center}
    \caption{Evasion techniques over TCP/IP protocols PART I.}
    \label{table:2.2}
    \begin{adjustbox}{width=\columnwidth,center}

    \begin{tabular}{lccccccccccccccccccccccc} 
      \toprule 
      \makecell{\multirow{4}{*}{\textbf{\makecell{APT \\ Campaign}}}} &  \multicolumn{8}{c}{\textbf{\cc\ Protocols}} & &\multicolumn{3}{c}{\textbf{\cc\ Tools}} & &\multicolumn{10}{c}{\textbf{\cc\ Evasion TTP}}         \\
        \cmidrule{2-9} \cmidrule{11-13} \cmidrule{15-24}
      &   \makecell{\multirow{2}{*}{\makecell{HTTP}}} &  \makecell{\multirow{2}{*}{\makecell{HTTPS}}}  & \makecell{\multirow{2}{*}{\makecell{DNS}}}  & \makecell{\multirow{2}{*}{\makecell{SMTP}}}  & \makecell{\multirow{2}{*}{\makecell{SOCKS5}}}  & \makecell{\multirow{2}{*}{\makecell{SMB}}}  & \makecell{\multirow{2}{*}{\makecell{FTP}}}  & \makecell{\multirow{2}{*}{\makecell{P2P}}}  && \makecell{\multirow{2}{*}{\makecell{RAT}}}  & \makecell{\multirow{2}{*}{\makecell{Bot.}}}  & \makecell{\multirow{2}{*}{\makecell{Backdoor}}}  & & \makecell{\multirow{2}{*}{\makecell{Proxy}}}  & \makecell{\multirow{2}{*}{\makecell{HTTP\\ Embedding}}} & \makecell{\multirow{2}{*}{\makecell{Obfuscation}}}  & \makecell{\multirow{2}{*}{\makecell{DNS \\tunneling}}} & \makecell{\multirow{2}{*}{\makecell{DGA}}}  & \makecell{\multirow{2}{*}{\makecell{Dynamic\\ DNS}}}& \makecell{\multirow{2}{*}{\makecell{Multi-Stage\\ Channels}}} & \makecell{\multirow{2}{*}{\makecell{Fallback \\Channels}}}& \makecell{\multirow{2}{*}{\makecell{ Multi-hop \\ Proxy}}} & \makecell{\multirow{2}{*}{\makecell{Multi-Layer \\Encryption}}}\\\\
      \midrule 
      \midrule 
      \\
     APT 1 & \checkmark & \checkmark & \checkmark & \checkmark & & &\checkmark& && \checkmark& & \checkmark& & && \checkmark &  &\checkmark & \checkmark & \checkmark&\checkmark&& \checkmark \\
     \\
APT 2 & \checkmark &  &  &  & & & &  && \checkmark& & \checkmark& & && \checkmark &  & &  &\checkmark&\checkmark&& \checkmark \\
\\
APT 3 & \checkmark & \checkmark & \checkmark  &  & \checkmark & & && & \checkmark& & \checkmark& & \checkmark&& \checkmark &  & &  &\checkmark&\checkmark&& \checkmark \\
\\
APT 10 & \checkmark & \checkmark & \checkmark &  & \checkmark & & \checkmark& && \checkmark& & \checkmark& & \checkmark& \checkmark& \checkmark &  & &  &\checkmark&\checkmark&& \checkmark  \\
\\
APT 12 & \checkmark &  &  &  & & & & & & \checkmark& & \checkmark& & \checkmark& \checkmark& \checkmark &  & &  &\checkmark&&&   \\
\\
APT 15 &\checkmark &  & \checkmark & & & & &  & & \checkmark& & \checkmark& & & & \checkmark &  & &  &&&& \\
\\
APT 16 &\checkmark & \checkmark &  && & &  &  & & \checkmark& & \checkmark& & \checkmark & & \checkmark &  & &  &\checkmark&&& \\
\\
APT 17 &\checkmark &  &  &  & & & & && & & \checkmark& &  & & \checkmark &  & &  &\checkmark&\checkmark&&\\
\\
APT 18 & \checkmark & \checkmark & \checkmark &  & & & & & & \checkmark& & \checkmark& & & & \checkmark &  & &  &\checkmark&\checkmark&& \checkmark\\
\\
APT 19 & \checkmark & \checkmark & \checkmark & & &\checkmark &\checkmark &\checkmark  & & \checkmark& & \checkmark& & & & \checkmark &  & &  &\checkmark&\checkmark&& \checkmark\\\\
APT 27 & \checkmark & \checkmark & \checkmark  &  & & &  && & \checkmark& & \checkmark& & & \checkmark& \checkmark &  & &  &\checkmark&\checkmark&& \checkmark \\\\
APT 28 & \checkmark & \checkmark & \checkmark  &  \checkmark& \checkmark & &&& & \checkmark& & \checkmark& & \checkmark& \checkmark& \checkmark & \checkmark  & \checkmark & \checkmark &\checkmark&\checkmark&\checkmark& \checkmark\\\\
APT 29 & \checkmark & \checkmark & \checkmark  &  \checkmark& &\checkmark & &\checkmark& & \checkmark& & \checkmark& & \checkmark& \checkmark& \checkmark & \checkmark  & \checkmark & \checkmark &\checkmark&\checkmark&\checkmark& \checkmark \\\\
APT 30 & \checkmark &  &  &  & & & & & & \checkmark& & \checkmark& & \checkmark& \checkmark& \checkmark &  & &  &\checkmark&&&  \\\\
APT 32  & \checkmark & \checkmark & \checkmark  &  & &\checkmark & &\checkmark& & \checkmark& & \checkmark& & \checkmark& \checkmark& \checkmark & \checkmark  & \checkmark & \checkmark &\checkmark&\checkmark&\checkmark& \checkmark\\\\
APT 33 &\checkmark & \checkmark &  & \checkmark& & &  &  & & \checkmark& & \checkmark& &  & & \checkmark &  & &  &\checkmark&&& \\\\
APT 34 & \checkmark &  & \checkmark  &  & & &\checkmark && & \checkmark& & \checkmark& & & \checkmark& \checkmark & \checkmark  & \checkmark & \checkmark &\checkmark&\checkmark&& \checkmark\\\\
APT 35  &\checkmark &  &  &  & & &\checkmark & \checkmark&& & \checkmark & & &  & & \checkmark &  & &  &&&&\\\\
APT 37  & \checkmark & \checkmark &   & \checkmark & &\checkmark & &\checkmark& & \checkmark& & \checkmark& & \checkmark& \checkmark& \checkmark & \checkmark  & \checkmark & \checkmark &\checkmark&\checkmark&\checkmark& \checkmark\\\\
admin@338 & \checkmark & \checkmark &  &  & \checkmark & & & && \checkmark& & \checkmark& & && \checkmark &  & &  &&&& \\\\
Blockbuster & \checkmark & \checkmark &   & \checkmark & & & & & & \checkmark& & \checkmark& & \checkmark& \checkmark& \checkmark &   &  &  &\checkmark&\checkmark&\checkmark& \checkmark\\\\
Cobalt Group & \checkmark & \checkmark & \checkmark  &  & &\checkmark & &\checkmark& & \checkmark& & \checkmark& & \checkmark& \checkmark& \checkmark & \checkmark  & \checkmark & \checkmark &\checkmark&\checkmark&\checkmark& \checkmark\\\\

      \bottomrule 
      \bottomrule 

    \end{tabular}
    \end{adjustbox}

  \end{center}
\end{table}
\end{landscape}

\begin{landscape}

\begin{table}[h!]
  \begin{center}
    \begin{adjustbox}{width=\columnwidth,center}

    \begin{tabular}{lccccccccccccccccccccccc} 
      \toprule 
      \makecell{\multirow{4}{*}{\textbf{\makecell{APT \\ Campaign}}}} &  \multicolumn{8}{c}{\textbf{\cc\ Protocols}} & &\multicolumn{3}{c}{\textbf{\cc\ Tools}} & &\multicolumn{10}{c}{\textbf{\cc\ Evasion TTP}}         \\
        \cmidrule{2-9} \cmidrule{11-13} \cmidrule{15-24}
      &   \makecell{\multirow{2}{*}{\makecell{HTTP}}} &  \makecell{\multirow{2}{*}{\makecell{HTTPS}}}  & \makecell{\multirow{2}{*}{\makecell{DNS}}}  & \makecell{\multirow{2}{*}{\makecell{SMTP}}}  & \makecell{\multirow{2}{*}{\makecell{SOCKS5}}}  & \makecell{\multirow{2}{*}{\makecell{SMB}}}  & \makecell{\multirow{2}{*}{\makecell{FTP}}}  & \makecell{\multirow{2}{*}{\makecell{P2P}}}  && \makecell{\multirow{2}{*}{\makecell{RAT}}}  & \makecell{\multirow{2}{*}{\makecell{Bot.}}}  & \makecell{\multirow{2}{*}{\makecell{Backdoor}}}  & & \makecell{\multirow{2}{*}{\makecell{Proxy}}}  & \makecell{\multirow{2}{*}{\makecell{HTTP\\ Embedding}}} & \makecell{\multirow{2}{*}{\makecell{Obfuscation}}}  & \makecell{\multirow{2}{*}{\makecell{DNS \\tunneling}}} & \makecell{\multirow{2}{*}{\makecell{DGA}}}  & \makecell{\multirow{2}{*}{\makecell{Dynamic\\ DNS}}}& \makecell{\multirow{2}{*}{\makecell{Multi-Stage\\ Channels}}} & \makecell{\multirow{2}{*}{\makecell{Fallback \\Channels}}}& \makecell{\multirow{2}{*}{\makecell{ Multi-hop \\ Proxy}}} & \makecell{\multirow{2}{*}{\makecell{Multi-Layer \\Encryption}}}\\\\
      \midrule 
      \midrule 
      \\

Dragonfly 2.0 &\checkmark & \checkmark &  &  & & &  & && & & & & \checkmark & & \checkmark &  & &  &&&&\\\\
Duqu 2.0 &\checkmark & \checkmark &  &  & & &  & && & & & & \checkmark & & \checkmark &  & &  &&&&\\\\
Elderwood &\checkmark & \checkmark &  &  & & & & &&\checkmark & & & & \checkmark & & \checkmark &  & &  &&&& \\\\
FIN 7 &\checkmark & \checkmark & \checkmark  &  & & & & &&\checkmark & & & &  & & \checkmark & \checkmark & &  & \checkmark &\checkmark&&\\\\
Leviathan & \checkmark & \checkmark & \checkmark  &  & &\checkmark & &\checkmark& & \checkmark& & \checkmark& & \checkmark& \checkmark& \checkmark & \checkmark  & \checkmark & \checkmark &\checkmark&\checkmark&\checkmark& \checkmark \\\\
Naikon APT & \checkmark &  &  &  &  & &\checkmark & & & \checkmark& & \checkmark& & \checkmark& \checkmark& \checkmark &  & &  &\checkmark&&&  \\\\
Patchwork  & \checkmark & \checkmark & &  & \checkmark & &\checkmark & && \checkmark& & \checkmark& & \checkmark& \checkmark& \checkmark &  & &  &\checkmark&\checkmark&& \checkmark \\\\
Sandworm & \checkmark &  & &  & \checkmark & & & && \checkmark& \checkmark & \checkmark& & & \checkmark& \checkmark &  & &  &\checkmark&\checkmark&& \checkmark\\\\
Strider & \checkmark & \checkmark & \checkmark & \checkmark & & & & && & & \checkmark& & && \checkmark &  & &\checkmark  & \checkmark&\checkmark&& \checkmark\\\\
Regin &\checkmark & \checkmark &  &\checkmark& & & &  & & & & \checkmark& & \checkmark & & \checkmark &  & &  &\checkmark&&&\\\\
Red October &\checkmark & \checkmark &  && & &\checkmark  &  & & & & \checkmark& &  &\checkmark & \checkmark &  & &  &\checkmark&&\checkmark&\\\\

      \bottomrule 
      \bottomrule 

    \end{tabular}
    \end{adjustbox}

  \end{center}
\end{table}
\end{landscape}

\section{Taxonomy of Network-based TTPs Used by APTs}\label{sec:apt-ttps}

\begin{figure}[t!]
\centering
\includegraphics[width=8.5cm,height=4.25cm]{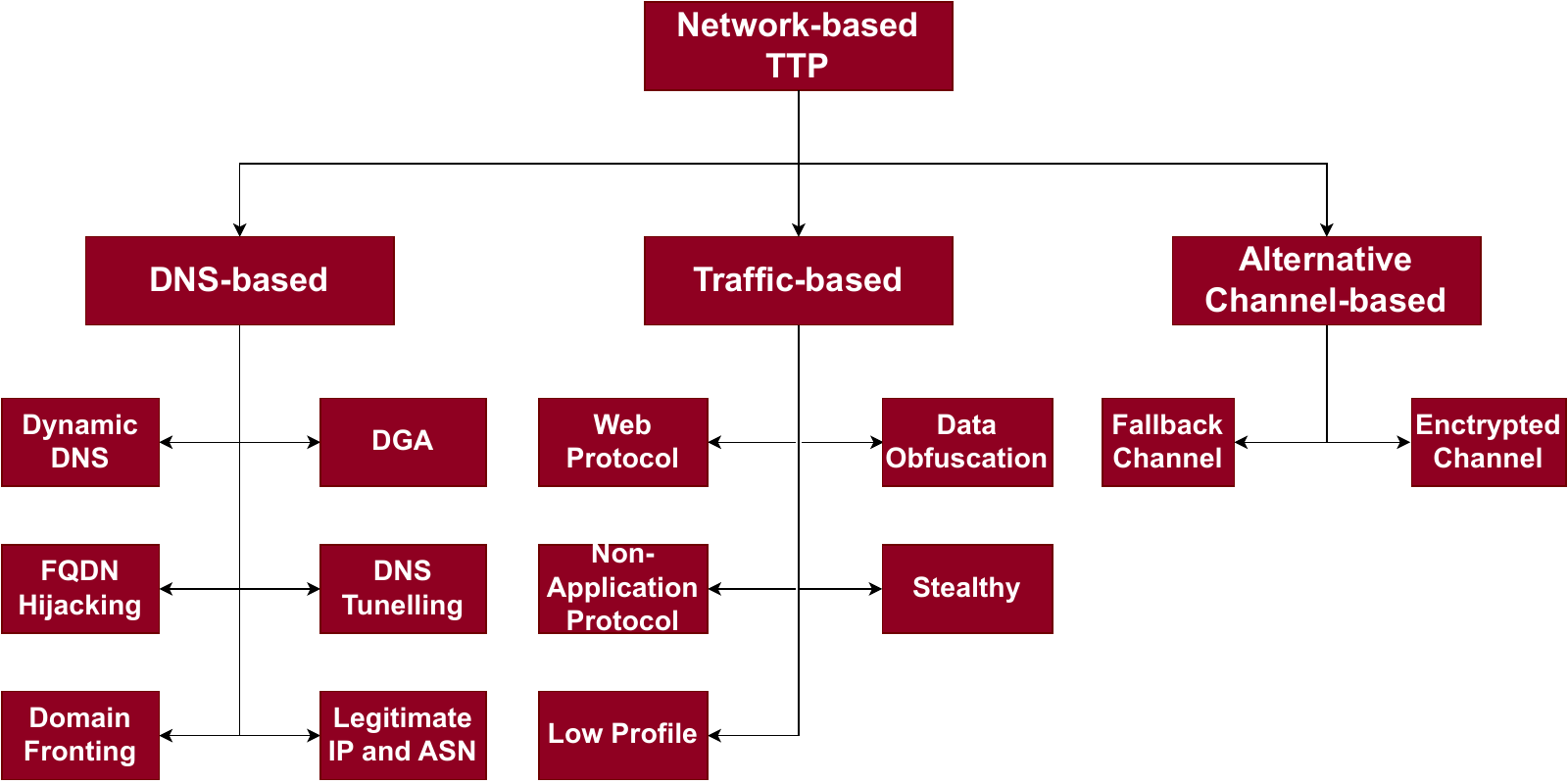}
\caption{Network-based TTPs taxonomy.}\label{fig:ttp_taxonomy}
\end{figure}

Figure \ref{fig:ttp_taxonomy} illustrates the taxonomy of the network-based TTPs based on our findings. The first level is categorized based on the common usage by users. For example, DNS-based is normally used for retrieving the DNS resources, such as the IP address of the remote server. 
In APTs, the possible techniques used to locate \cc\ servers include Dynamic DNS, FQDN Hijacking, Domain Fronting, and stealthy DGA. However, we include DNS tunneling in this category as it appears to a SOC analyst to be DNS packets. The second category relates to traffic-based techniques. We limit this category to our findings based on our datasets with additional information based on the technical reports. This includes web protocol, data obfuscation, non-application protocol, DNS over HTTPS (DoH), low profile, and stealthy behavior. DoH remains in this category since it uses HTTPS and is not visible to a SOC analyst unless they have the key to decipher the packets at the web proxy. 
The final category is the possible channels that can be used during an attack. The two techniques frequently used for APTs are the encrypted and fallback channels. We keep these two techniques under the channel-based category because they can act as the backbone for other TTPs, and enhance the difficulty of detecting them.

\section{DNS-based TTPs} \label{sec:dnsttps}

The first line of defense is to stop the APT traffic before it begins, which relies on detecting malicious domains. However, Malicious domain detectors can be evaded using DNS-based TTPs. In this section, we discuss the TTPs related to DNS reported in Figure \ref{fig:ttp_taxonomy}.

\subsection{IP Addresses and ASN}
Analysts in \cite{aptNaikonB} introduce an analysis of Naikon APT and observe that 99\% of the campaign's domains use two IP addresses, and it can reach up to 51 IPs. This suggests that counting the IP addresses used for a single domain over a time window might be useful. However, if we consider the autonomous system number (ASN) for a given IP address, then we could identify a malicious cluster. For example, in Naikon APT, six clusters are identified in South Korea (Seoul), China (Kunming and Jinma), Thailand (Bangkok), and the USA (Denver), with 22 city DNS resolutions referring to Kunming in China. \cite{apt27a} reports that threat actors of APT 27 frequently change the IP address of \cc\ domain, but within the same subnet. 
\begin{figure*}[h]
\centering
\includegraphics[scale=0.34]{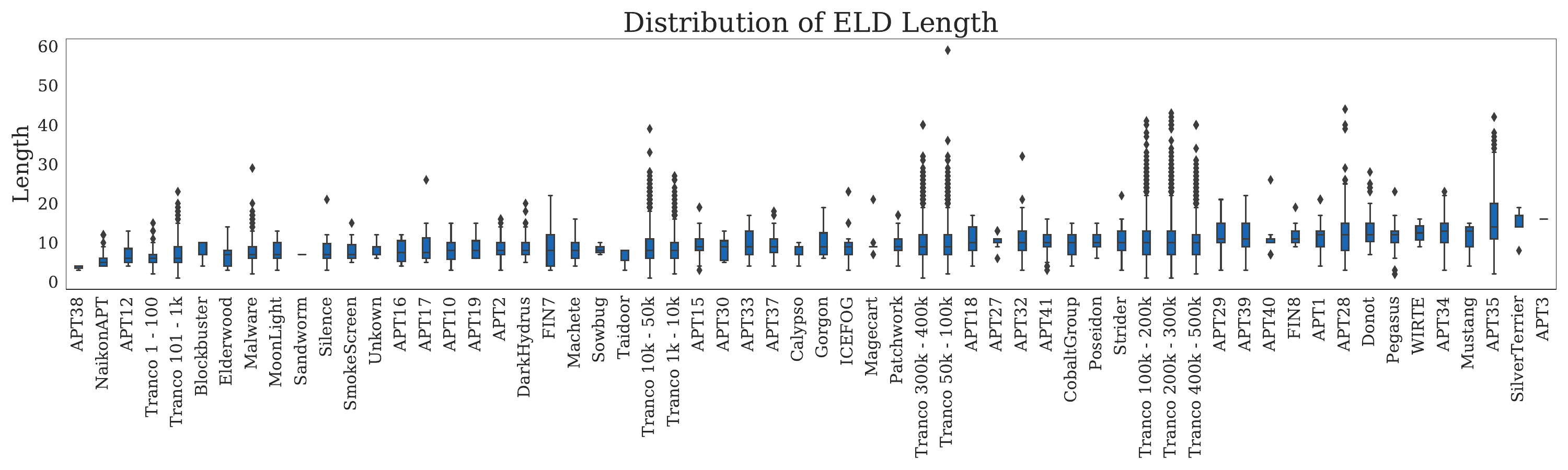}
\caption{Campaigns are sorted by median length. APT and legitimate domain campaigns are interleaved. The upper and lower caps refer to the 90th and 10th percentile, whereas the upper and lower sides for each box represent the 75th and 25th percentile. Finally, the mid-line inside boxes are median values, and diamonds refer to outliers.} \label{figure:APT_length}
\centering
\end{figure*}

\subsection{Hijacked FQDNs}
 While the classic DNS spoofing attack targets the DNS recursive resolver's cache to forward the request to a malicious IP, FQDN hijacking targets the legitimate web server. FQDN Hijacking is the process of hijacking an FQDN with a qualified registered name and legitimate use \cite{apt1}. First, threat actors penetrate the web server hosting a legitimate website. Then, they implant a backdoor on their web pages and send spear-fishing emails with a link to the victim. After the victim is hacked, the backdoor is preconfigured to communicate to \cc\ using the hijacked FQDN. For the network defenses, the hijacked FQDN is a legitimate one, and the NIDS or SOC team cannot discover that the legitimate domain becomes a malicious one. We notice that APT 1 \cite{apt1}, APT 27 \cite{apt27b} and APT 34 \cite{apt34b} frequently use such a technique for their \cc\ operations.

\subsection{DNS tunneling}

APT 32, APT 34, Cobalt Group, and Strider campaigns use DNS tunneling. A threat actor might embed restricted protocols inside a legitimate one, such as DNS, HTTP, and SSH. The idea of DNS tunneling is to communicate to \cc\ server via DNS query  \cite{born2010detecting}.  For instance, APT 32, a.k.a OceanLotus, uses DNS tunneling to avoid NIDS and to bypass Firewalls \cite{apt32c}. The destination IPs are legitimate DNS servers, i.e., Google and OpenDNS \cite{apt32c}. However, the malicious domain is embedded in the packet, which will be unpacked at some intermediate points to forward the traffic accordingly \cite{qi2013bigram}. For instance, Pisloader sends a beacon periodically, with setting flags including response, recursion desired, and recursion, and in case any additional flags are set, the packet will be discarded \cite{apt18d}. The obfuscated based64 payloads are attached on the same string with the \cc\ server domain in a 4-bytes length \cite{apt18d}. The \cc\ server responds in the TXT record of the DNS packet with the same settings \cite{apt18d}.

\subsection{Dynamic DNS}
Since 2016, Dynamic DNS with third-party services has been essential for several APTs \cite{apt10a}, including APT 1 \cite{apt1}, APT 2 \cite{apt2a}, APT 10 \cite{apt10b}, APT 15 \cite{apt15a}, APT 18 \cite{apt18a}, Leviathan \cite{apt40a} and Naikon \cite{aptNaikonA}. APT actors frequently register new subdomains under sharing zones with other clients. APT 1, for instance, registered hundreds of subdomains over the years and frequently changed the resolution of FQDN to new IP addresses of \cc\ servers by logging into the service via a web-based interface and updating instantly \cite{apt1}. NIDS and the security team view the malicious FQDN using Dynamic DNS as a domain from a public service, which makes it difficult to classify it as malicious.

\begin{table}[h]

    \small
    \begin{tabular}{p{2cm}p{5.2cm}} 
      \toprule 
         Set  & ELD (.TLD) \\
       \midrule 
       \midrule 

        
       \multicolumn{2}{c}{\textbf{I. Bitsquatting}}\\
        APT 35& telagram ([.]net) \\

       Legitimate& telegram ([.]org)\\\\

              \multicolumn{2}{c}{\textbf{II. DGA-Like Alexa-based }}\\

        APT 34	&	egoogle ([.]org)	\\
        Legitimate &	google ([.]org)\\\\

              \multicolumn{2}{c}{\textbf{IV. DGA-Like Dictionary-based} }\\

       ICEFOG&	sportsnewsa ([.]net)\\ 
        Legitimate  & sportsmansoutdoorsuperstore ([.]com)  \\

              \multicolumn{2}{c}{\textbf{V. DGA-Like Random Character} }\\

         APT 28  & 	message-id8665213 ([.]com) \\
        Legitimate &  xn----8sbkahkuskl1n ([.]com) 
  \\\\

        \multicolumn{2}{c}{\textbf{VI. Dynamic DNS With Third Party} }\\

         APT 15 	&ensun ([.]dyndns[.]org)\\
       Legitimate &	erogamescape ([.]dyndns[.]org)\\\\

              \multicolumn{2}{c}{\textbf{VII. Technical Process} }\\

         APT 28 &	accoounts-google ([.]com)	\\ 
         Legitimate &	googleaccountlogin ([.]com)
 \\\\

              \multicolumn{2}{c}{\textbf{VII. Typosquatting} }\\

        APT 34	& miedafire ([.]com)	\\\\

       \multicolumn{2}{c}{\textbf{II. TLD Squatting}}\\
       WIRTE & microsoft ([.]store) \\
        Legitimate  & microsoft ([.]com)  \\\\
              \multicolumn{2}{c}{\textbf{IX. Phishing} }\\

       Patchwork	&yahoomail ([.]pw|[.]support)	\\
         Legitimate &	(mail[.]) yahoo ([.]com)\\

      \bottomrule 

    \end{tabular}
\raggedright
    \caption{Examples of APT ELDs across different evasion techniques.}     \label{table:extended_examples_seg}
\end{table}
\subsection{Stealthy Domain Generation Algorithms (DGA)}\label{ch3:sec:dga}

DGAs generate a large number of pseudo-random variations of domain names using algorithms based on arithmetic, hash functions, wordlists, and permutations~\cite{plohmann2016comprehensive}. 
%
Normally, DGA is used as a backup plan for the botnets to communicate back to a master bot \cite{sood2016taxonomy}. Zeus and Cryptolocker are pretty famous for using binary DGAs modules attached to the malware to algorithmically generate domains (AGDs) until \cc\ channel is established \cite{sood2016taxonomy}. Once the host is infected by malware with the DGA module, the module will start sending a random DNS query with a seed value known upfront by the threat actor. The communication will then be initiated with the \cc\ server if the DNS query matches a registered domain. 
In the APT context, APT 28 and APT 34 are the main examples of campaigns using DGA. Backdoor uses DGA to create more domains upon the request of \cc\ servers in hours and destroy it within 24 hours \cite{fu2017stealthy}. 
These make it hard to predict what domain names will be used, and even when the algorithm is successfully reverse-engineered, take-downs are expensive as only a small fraction of a large number of generated domains is effectively registered by an adversary.

\subsection{Domain Fronting and Multi-hop Proxy} \label{sec:Multi_hop_Proxy}
\begin{figure*}[!t]
\centering
\includegraphics[width=17cm,height=4cm]{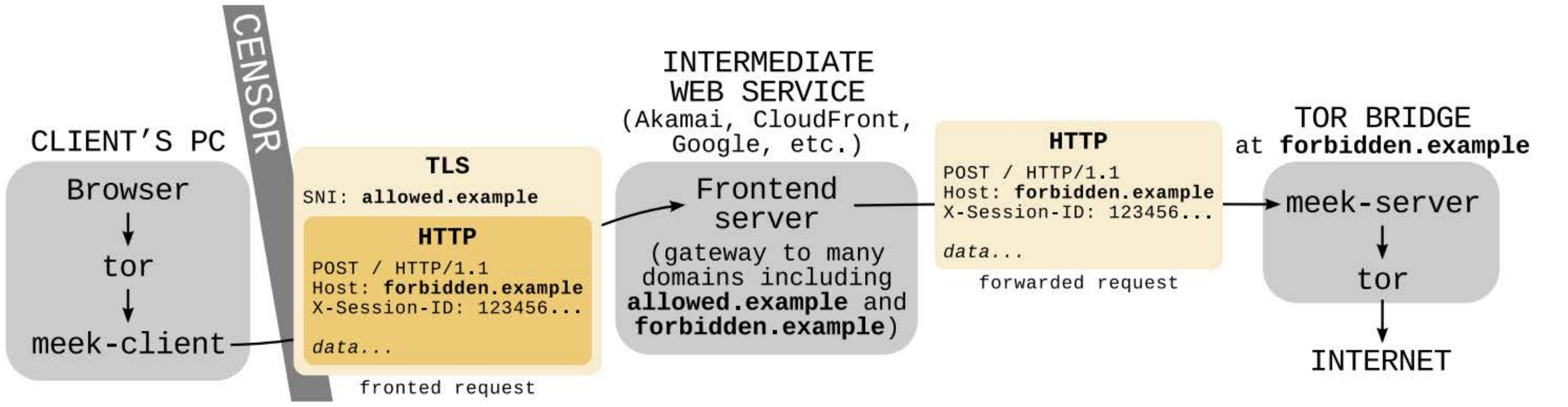}
\caption{An example of domain fronting \cite{fifield2015blocking}.}\label{fig:domain_fronting}
\end{figure*}

Domain fronting is an evasive technique that conceals the location of the remote \cc\ server  \cite{fifield2015blocking}. The backdoor implanted in a victim machine pretends to communicate with a legitimate host over HTTPS. The legitimate host is only an intermediate step to communicate with the true destination. This chain may include multiple hops. The domain outside of the HTTPS request is a legitimate one, while the inside the HTTPS is the true destination, which is encrypted (Figure \ref{fig:domain_fronting}).  

We observe that APT 29 \cite{apt29a} and Blockbuster \cite{LazarusGroupC} use domain fronting and multi-hop proxy. 
APT 29 connects to the onion router (TOR) with the domain fronting technique, i.e., a domain included in the TLS header refers to Google services. The destination domain (TOR multiple hops) is hidden in the HTTP header until it connects back to a threat actor \cite{apt29d}. To update the FQDN for newer \cc\ servers, the Duke malware may receive the IP address and exfiltrate over Twitter and Microsoft One Drive \cite{apt29b}. In addition, the TOR tunnel provides remote access to the victim by using Server Message Block (SMB) and NetBIOS. Throughout this operation, the HTTP POST header appears to the NIDS as a legitimate URL with unsuspicious parameters \cite{apt29d}.

\section{Traffic-based TTPs}\label{sec:trafficttps}
After APT operators successfully evade the defenses and locate the \cc\ server, they deploy several traffic-based TTPs to continue undetected.  This section discusses the TTPs related to the traffic itself. 

\subsection{Web Protocol}
In order to evade detection and network filtering, adversaries may communicate by utilizing application layer protocols typical of online traffic. The protocol data exchanged between a client and server will contain instructions for the remote system and, typically, the responses to those instructions.
Internet communication protocols like HTTP(S) \cite{apt2a} and WebSocket \cite{BrazKing} may be widely used. Many different kinds of information can be hidden in the various fields and headers of an HTTP(S) packet. An attacker can utilize this to communicate with target systems under their control while appearing to be legitimate network traffic.

\begin{figure}[h]
\centering
\includegraphics[width=8.5cm,height=3.5cm]{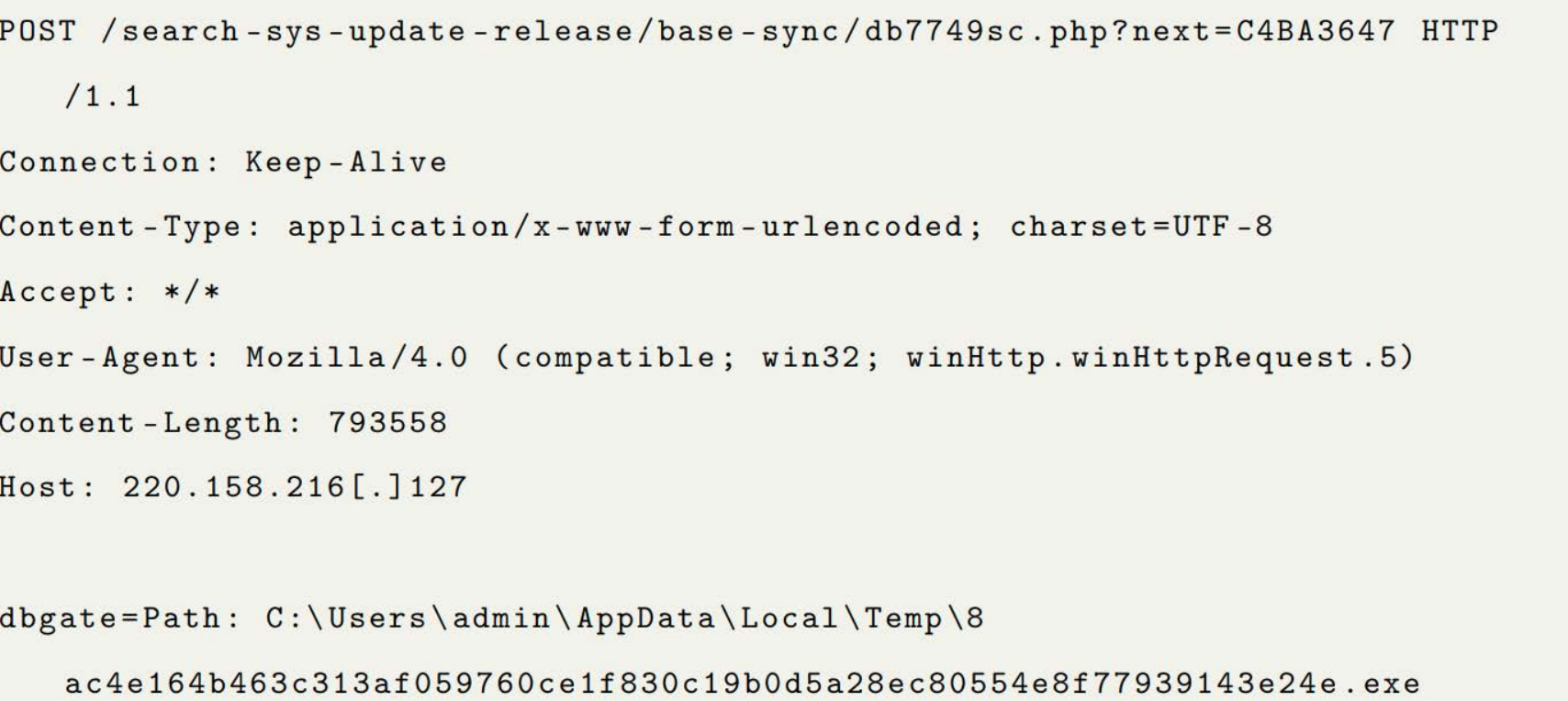}
\caption{ Zebrocy pushes collected information of the infected host using the POST method with WinHTTP library.}\label{fig:listing}
\end{figure}

In our dataset described in Section \ref{sec:meth}, APT28 uses a malware called Zebrocy. The malware has been observed spreading through maliciously produced Word documents with malicious macros and Dynamic Data Exchange (DDE)-based via social engineering or attachments targeting governments \cite{zebrocy_apt28}.
In our analysis, we confirm that the information of the infected hosts is collected and then sent back to 220.128.216[.]127 using HTTP with the POST method. In Figure \ref{fig:listing}, we extract the HTTP header and identify the usage of WinHTTP, a library often used as a web crawler or bot.  

\subsection{Data Obfuscation Through Protocol Impersonation}
\begin{figure}[h]
\centering
\includegraphics[width=8.5cm,height=3.5cm]{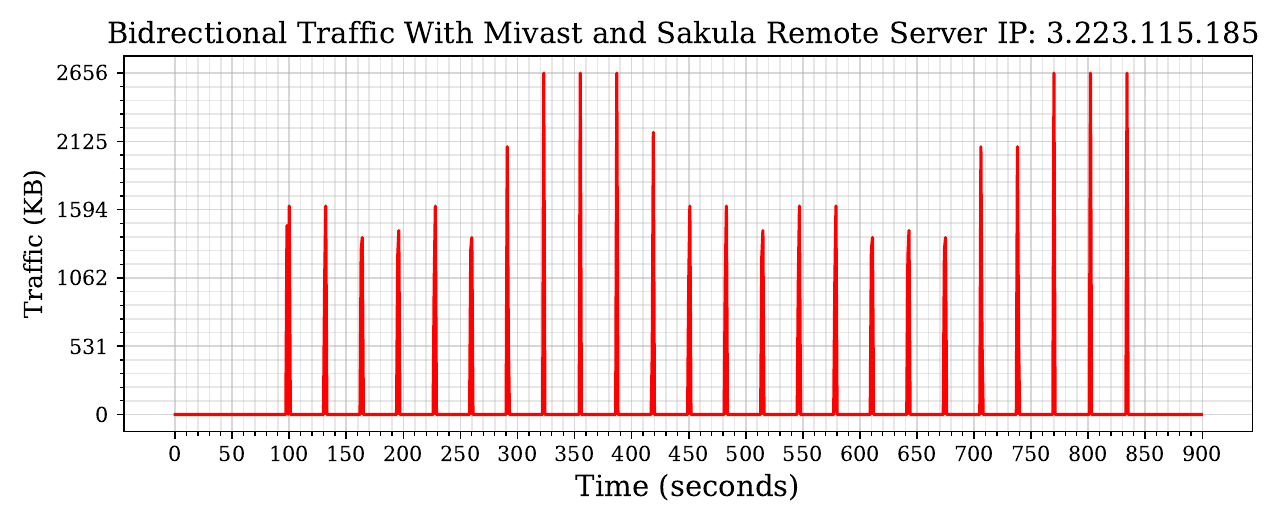}
\caption{Mivast and Sakula malware gather information and send it back to the operator periodically through data obfuscation through protocol impersonation.}\label{fig:data_obfuscation_profiling}
\end{figure}

A common tactic used by adversaries to prevent analysis and conceal \cc\ activity is to forge protocol or web service communication. In order to conceal their \cc\ traffic, adversaries may masquerade as genuine protocols or online services.
To trick security tools into thinking the following traffic is encrypted with SSL/TLS or to make their traffic appear to originate from a trusted source, attackers can spoof an SSL/TLS handshake. For example, A false TLS handshake is generated by Bankshot using a public certificate to camouflage \cc\ network communication \cite{Bankshot}. Cobalt Strike simulates HTTP while concealing malicious data by adding it to the URI, hiding it in headers, parameters, or transaction bodies, or using a combination of these methods \cite{CobaltGroupA}. Okrum uses a protocol that is quite similar to HTTP for \cc\ communication, but it conceals the actual messages in the Cookie and Set-Cookie headers of the HTTP requests that it makes \cite{OKRUM}. In our dataset, we commonly observe protocol impersonation through fake TLS over HTTP. In Figure \ref{fig:data_obfuscation_profiling}, we observe beaconing Mivast and Sakula malware adopt such a technique while the traffic is sent as impulses in periodic time slots. 

However, the protocol impersonation technique may include customized protocol. For instance, SolarWinds use the Orion Improvement Programme (OIP), where their users can provide feedback on the quality, functionality, and usefulness of their products in order to guide future development. Data about errors and other unusual occurrences in the system is also compiled \cite{SolarWinds_OIP}. Therefore, The SUNBURST Backdoor is able to blend in with legal SolarWinds activity because it disguises its network traffic as the OIP protocol and keeps its reconnaissance results within legitimate plugin configuration files \cite{SolarWinds_mandiant}.

\subsection{Non-Application Protocol}
An adversary may use a non-application layer protocol for communication between a host and a \cc\ server or between infected hosts inside a network \cite{apt41brown}. Numerous protocols can be considered. Network layer protocols like Internet Control Message Protocol (ICMP), transport layer protocols like User Datagram Protocol (UDP), session layer protocols like Socket Secure (SOCKS), and redirected/tunnelled protocols like Serial over LAN (SOL) are all used as examples.

\begin{figure}[h]
\centering
\includegraphics[width=8.5cm,height=3.5cm]{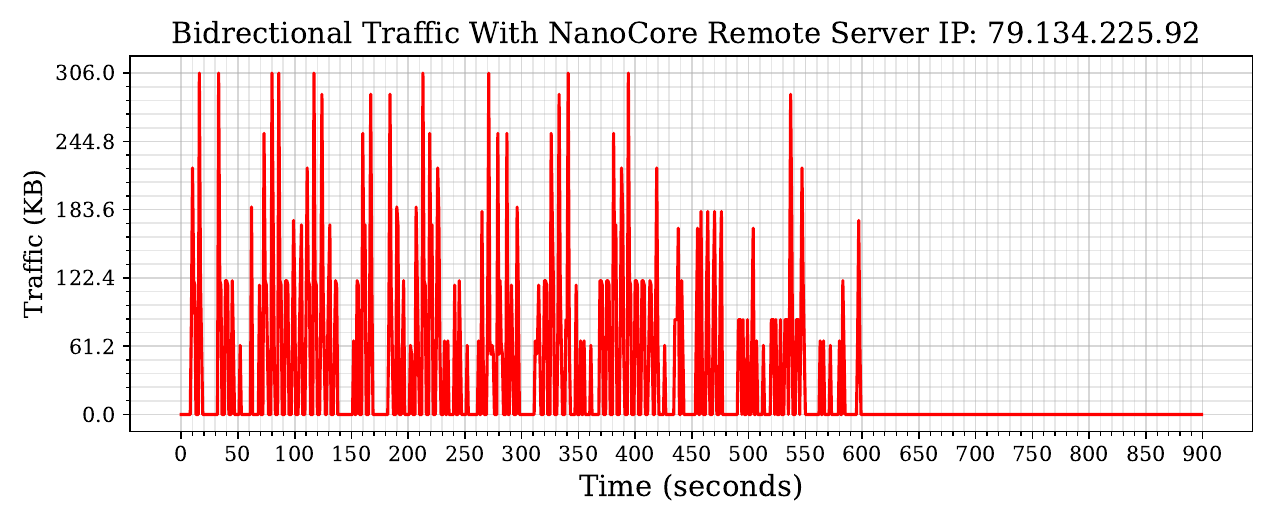}
\caption{NanoCore executes remote commands and collects information over Raw TCP.}\label{fig:non_app_profiling}
\end{figure}

Host-to-host communication via ICMP is one example. Since ICMP is included in the Internet Protocol Suite, all IP-compatible hosts must support it. However, unlike TCP and UDP, it is not as widely monitored and could be used by attackers looking to conceal conversations \cite{CISCO_IOS}.
In Figure \ref{fig:non_app_profiling}, we show an APT operator using NanoCore interacting with the victim using raw TCP only. 

\subsection{Stealthy Behavior}
\begin{figure}[h]
\centering
\includegraphics[width=8.5cm,height=7.5cm]{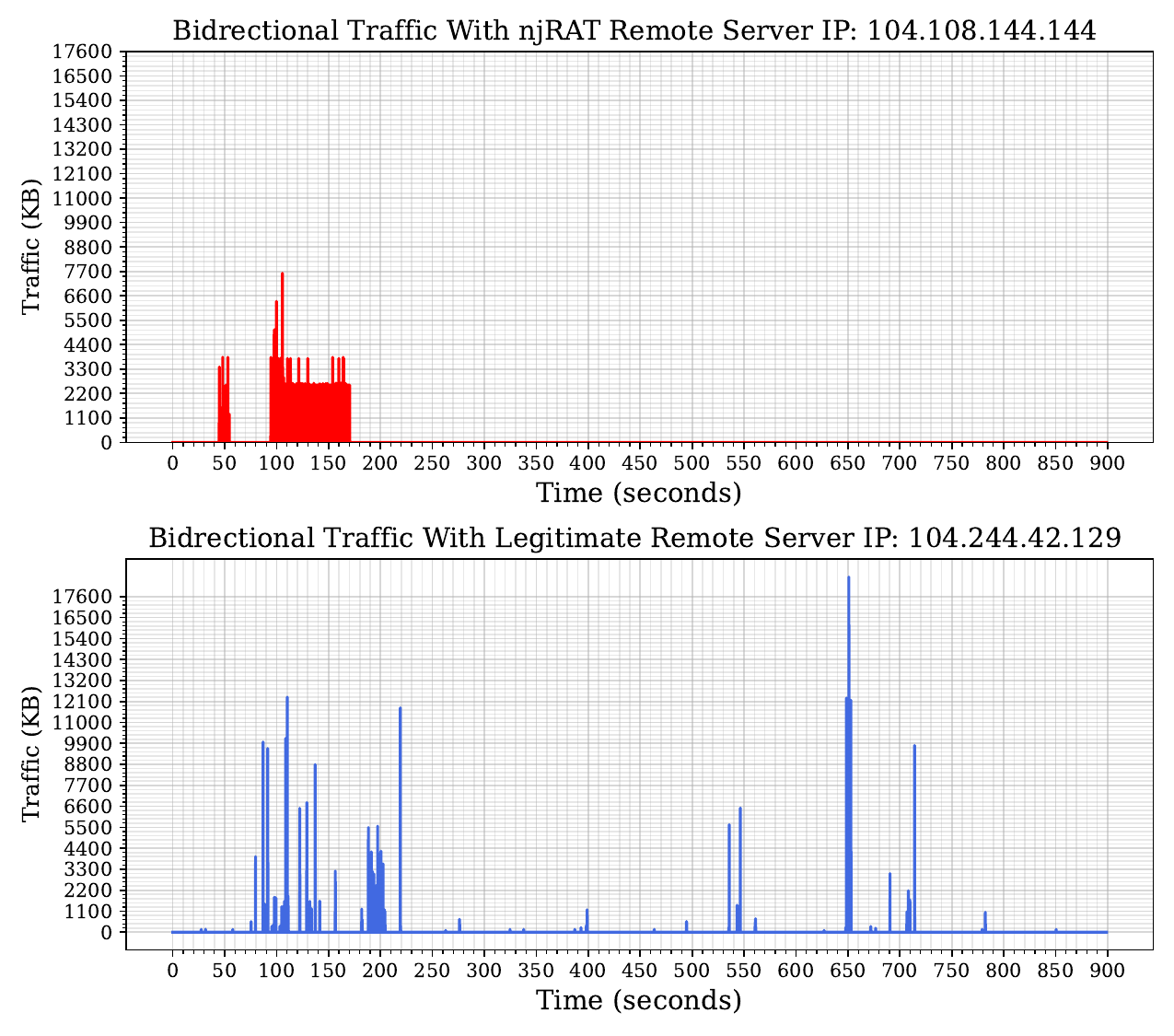}
\caption{Stealthy malicious APT StringPity compared legitimate behavior.}\label{fig:stealthy_profiling}
\end{figure}

Many reports blend stealthy behavior with a low profile. However, our findings show that stealthy behavior should be categorized as a different concept. Stealthy behavior is normally used for uploading a malicious payload. 
That means some packets are sent/received in a burst within a short period. For example, based on our analysis, njRAT exchanges data with the APT \cc\ within 76.53 seconds to drop the payload and exfiltrate initial information. In Figure \ref{fig:stealthy_profiling}, we show the difference between the stealthy behavior and the legitimate one. We notice that the volume of the burst does not exceed the legitimate threshold. 
In our example, the maximum magnitude of the burst for a single second is 7,614, compared to 18,630 KB for the legitimate one. This means the stealthy packets consume only 40.87\% of the legitimate threshold. However, we can see that the njRAT stealthy packets exchange data nearly without pausing after executing remote commands at the beginning. Once the mission is done, no packets are exchanged in the same time window. 

\subsection{Low Profile}
Most of the reports produced by governments and security vendors emphasize that the APT traffic normally operates at a low profile. That means an adversary is aware of the typical traffic volume in an enterprise network. An adversary may guess based on the number of hosts connected in the internal network. During the initial compromise stage, an adversary may gather the information around the network to minimize their traffic at a low profile. 

\begin{figure}[h]
\centering
\includegraphics[width=8.5cm,height=3.5cm]{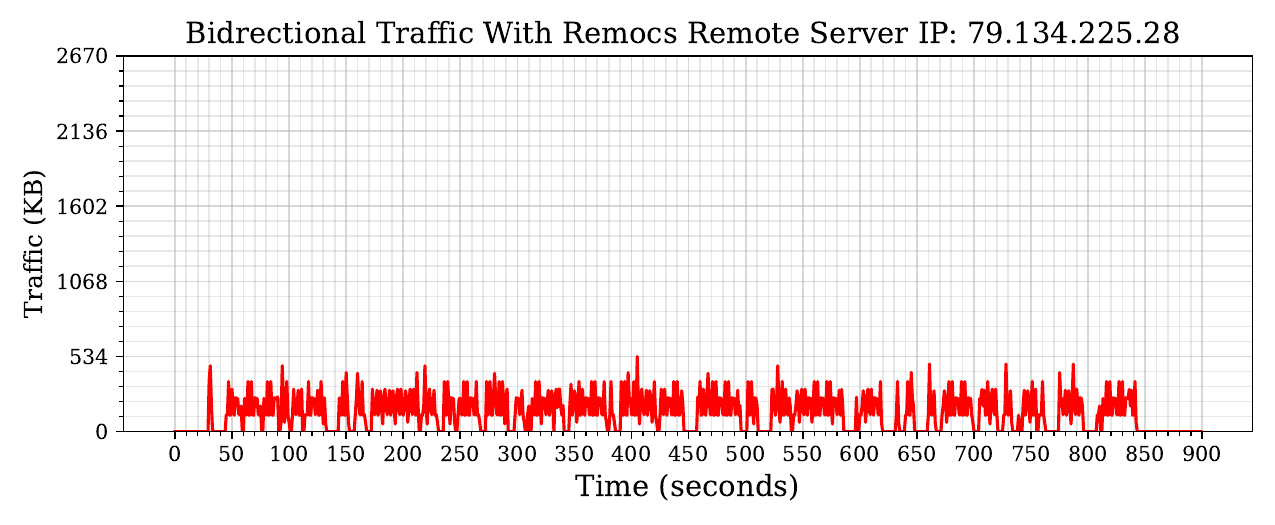}
\caption{Remcos transfers the collected data of the infected machine.}\label{fig:low_profiling}
\end{figure}

We can see the low magnitude of the malicious traffic for most TTPs in the figures presented in this section. However, let us take another example that stays in low profile mode while it does not use any previous traffic-based TTPs other than the non-application TTP. Figure \ref{fig:low_profiling} shows the behavior of Remcos. Remcos is a highly developed RAT that allows remote control and monitoring of any Windows machine running XP or later. Data is collected and transmitted to \cc\ servers, including OS version, user access, and processor revision number and architecture \cite{remcosRAT}. 

\section{Alternative Channel-based TTPs}
As described before, APT campaigns establish alternative channels for either persistence or NIDS obstruction. In this section, we describe the most common techniques used by most APTs. 

\subsection{Encrypted Channel} \label{sec:encrypted_channel}
It is well-known that several malware families use encryption for their communication. However, a multi-level encryption channel can be adopted in the APT settings. A common technique is to use Base64 encoding to disguise the remote commands using another encoding scheme such as UTF, ASCII or, simply, rotate by 13 places (ROT13). Next, the output is XORed the plaintext with the pad, and encrypt the output using symmetric encryption algorithms such as block ciphers (e.g. AES) or stream ciphers (e.g. RC4).  Later, when the traffic is carried over a trusted VPN, SOCK5, or TLS, the local network cannot identify the real information even after deciphering the connection. In this way, the local NIDS can be evaded as it sees the traffic came from public encrypted services such as GitHub, Dropbox, GoogleDrive, or any account on the social media platform.

\subsection{Fallback Channel} \label{sec:fallback_channel}
APT operators establish alternative channels with different IP addresses for two purposes. 
First, maintain persistence by providing a backup channel if one is detected and blocked. 
Second, stay undetected by following a divide-and-conquer strategy, which is the main reason for using fallback channels as a backbone to enable other TTPs deployment. We focus on this goal for the rest of our discussion and analysis here. 
As we know that APT operators carry out their different operations, such as payload download, controlling the victims with remote commands, crawling the information of the infected network, and exfiltrating data.

In practice, it is nearly impossible, with the current defenses in highly secure environments, to accomplish these operations without being detected.
Therefore, APT campaigns establish many remote servers to communicate with and split the communications among them to appear several legitimate behaviors from the enterprise point of view. For example, the first channel can be dedicated to downloading an internal reconnaissance tool by a combination of stealthy and web protocol TTPs. In addition, a new domain is sent using one of the DNS-based TTPs to evade malicious traffic and domain detectors.

\begin{figure}[t!]
\centering
\includegraphics[width=8.5cm,height=12cm]{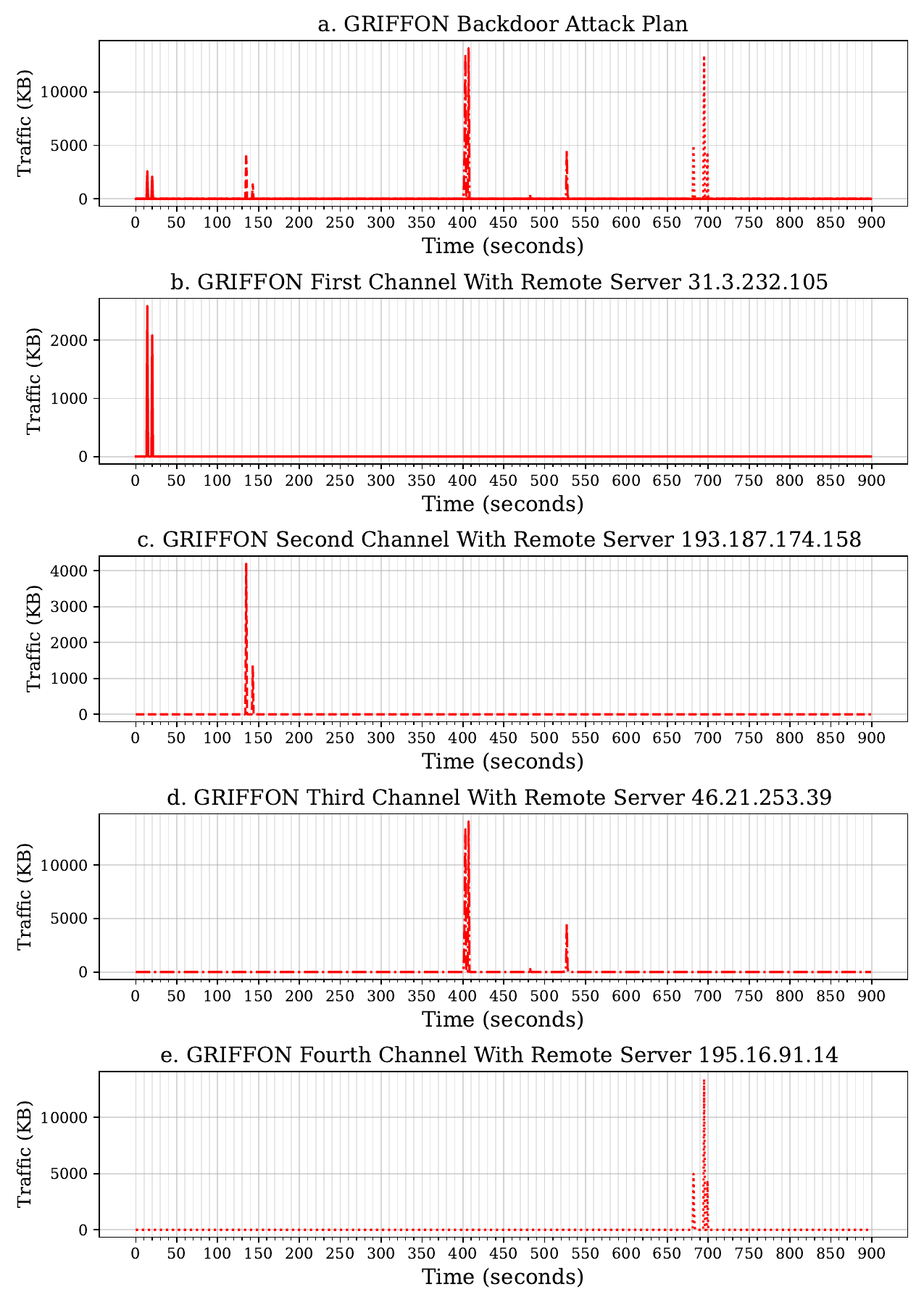}
\caption{GRIFFON divides its communication with its \cc\ over four channels.  }\label{fig:fallback_profiling}
\end{figure}

In Figure \ref{fig:fallback_profiling}.a, we plot the planned traffic to be exchanged with \cc\ server. However, Figure \ref{fig:fallback_profiling}.b-e depicts the actual behavior, which is split over four \cc\ servers during the same 15 minutes. 
Now, the first channel carries out the remote commands and retrieves initial internal reconnaissance information of the infected network. Just before being idle (Figure \ref{fig:fallback_profiling}.a), another domain is sent for the next channel. With resolving the domain name sent during the first channel, a second channel starts to communicate to another \cc\ server. The second channel continues retrieving more internal information with another domain sent for the next channel. Now, the third channel transfers a payload with two large impulses. Further tools are also sent over the last channel. Finally, APT operators prepare the tools needed for further tasks. For example, they can keep a logger and transfer the usernames and passwords, or they can continue with the lateral movement stage to control another victim in the same network. 

\section{Discussion} \label{sec:diss}
In this section, we summarize and discuss our findings based on the investigation of 118 technical reports and the analysis of the datasets introduced in Section \ref{sec:meth}.
We notice that over the past 22 years, spearphishing links have been popular to deliver the backdoors until now, which emphasizes the importance of developing a detective approach to defending against malicious domains. Spearphishing attachments are also popular and can break the current host-based intrusion detection, which is out of the paper's scope, and we leave it for future work.

\begin{figure*}[h]

\centering
\subfloat[\centering Most popular delivery methods.]{{\includegraphics[width=7.5cm,height=6cm]{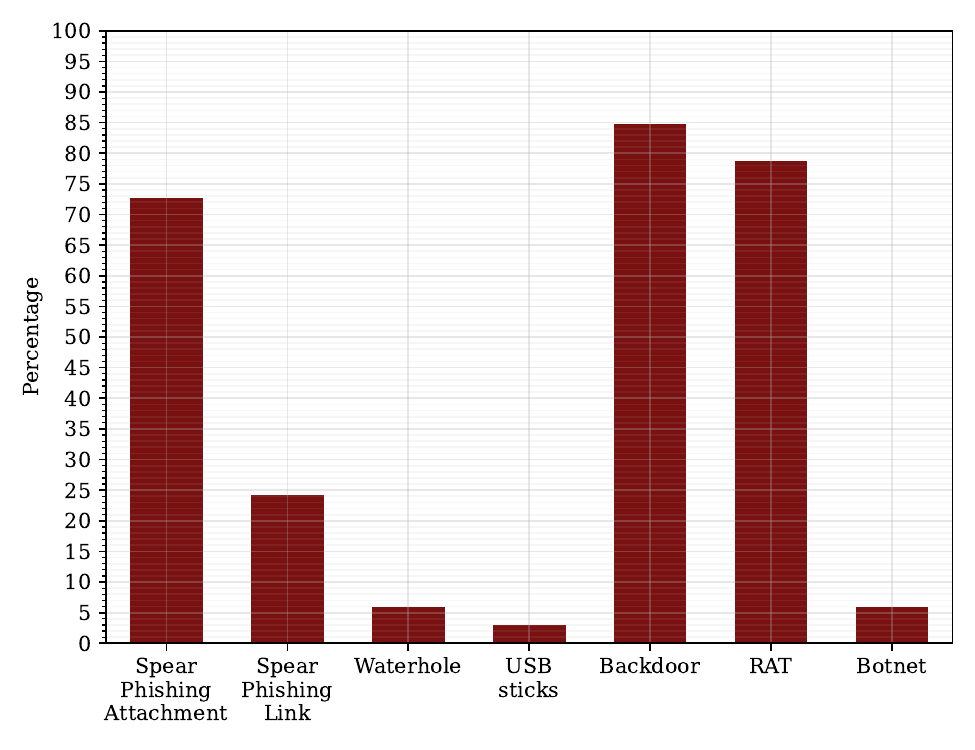} }}%
\qquad
\subfloat[\centering Most popular protocols usage.]{{\includegraphics[width=7.5cm,height=6cm]{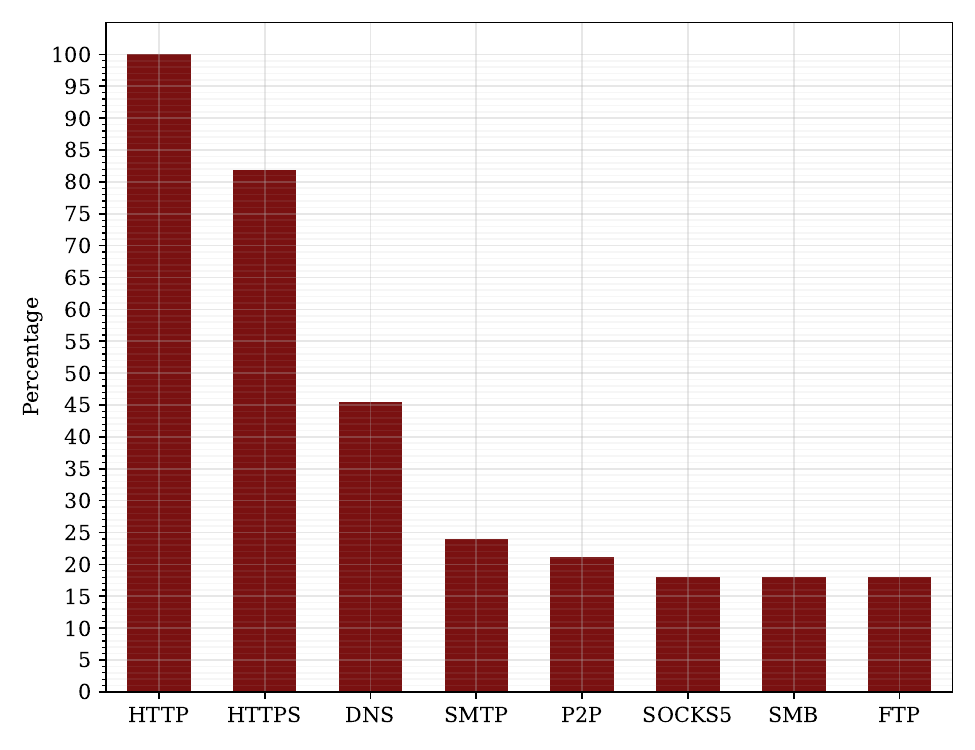} }}%
\qquad
\subfloat[\centering Top TTPs usage.]{{\includegraphics[width=10.5cm,height=6cm]{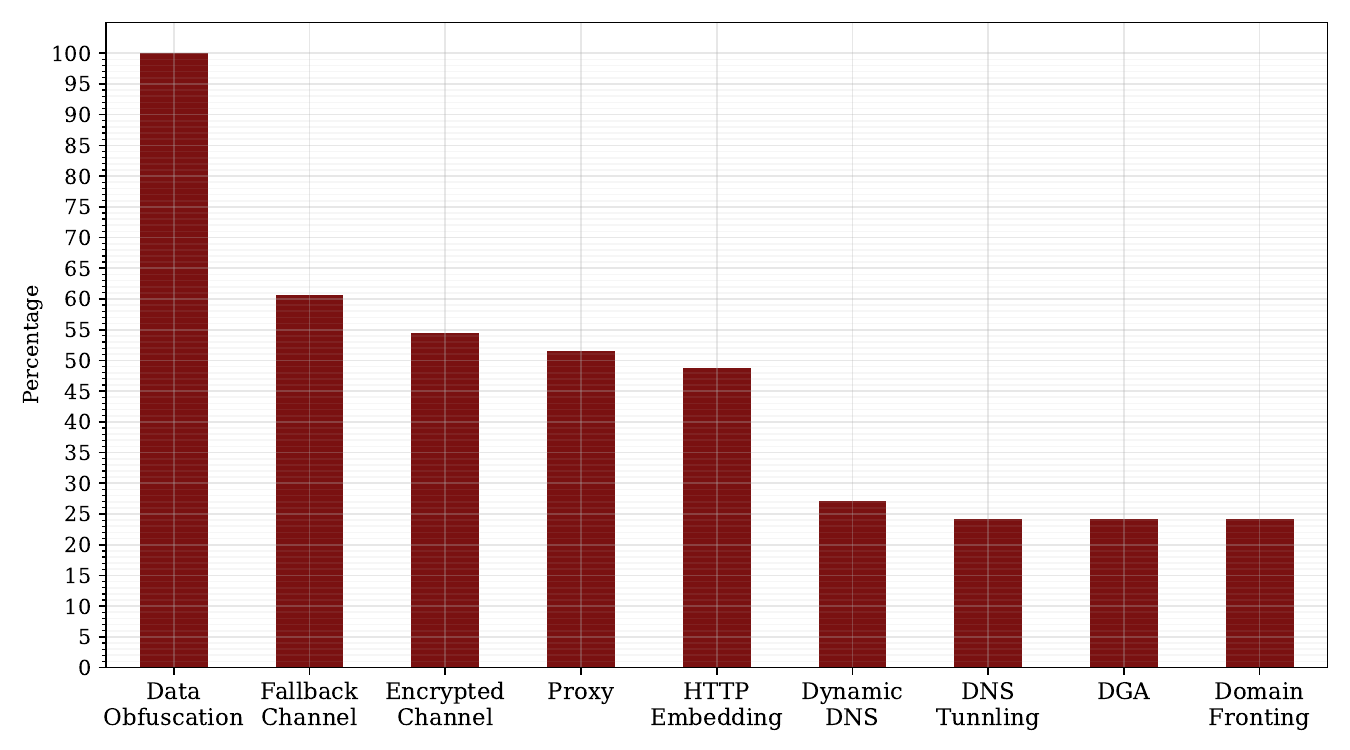} }}%
\caption{Investigation summary based on 33 APT campaigns.}\label{fig:profiling_results}
    
\end{figure*}

In Section \ref{Section: Analysis of APT}, we discuss if the APTs are a single malware or an arsenal of malware. We reported the backdoors for each campaign to confirm that APTs used a collection of malware, for instance, up to 26 different malware families used by APT 1. Some of these backdoors are communicating with the remote \cc\ server. There is a chance to find if an organization is under an APT campaign attack if we are able to detect the malicious traffic by such malware. We find that 84.8\% of APT campaigns use malware that is described as backdoors and 78.7\% are RAT while only 6\% are botnets as depicted in Figure \ref{fig:profiling_results}.a. That clarifies that APTs rarely use botnets for their attack and rely heavily on customized backdoors and RATs. 

Since our scope in this paper is to defend against APTs using network data, we limit the rest of the discussion to our findings related to the popular protocols and TTPs used to communicate to \cc\ server.

Figure \ref{fig:profiling_results}.b shows that all APT campaigns have continued using HTTP since 2001 to evade NIDS detection, emphasising that we need to carefully detect malicious connections that use HTTP protocol. That does not mean every single connection will use HTTP protocol, but it means at least every APT campaign will use HTTP once at a time. 81\% of APT campaigns rely on HTTPS to bypass NIDS that rely on HTTP plaintext features. Since HTTP protocol normally starts with DNS domain resolution, 45\% of APTs use DNS protocols, while the others use HTTP with a preconfigured IP address.
Also, 24.2\% of APT campaigns deliver their malware using spearphishing links, which we should also consider to detect malicious URLs. SMTP, P2P, SOCKS5, SMB and FTP come next with 24\%, 21.2\%, 18.1\%, 18.1\%, and 18\%, respectively.

To design a reliable approach to detect APTs, we study their usage of TTPs that enable them to bypass NIDS. Figure \ref{fig:profiling_results}.c shows that APTs typically adopt data obfuscation through protocol impersonation. We show an example of Mivast and Skula malware in Figure \ref{fig:data_obfuscation_profiling} impersonating legitimate HTTPS. The next popular TTP is the fallback channel to split the traffic volume over multiple \cc\ servers, which is used by 60.6\% of APT campaigns. This percentage is increasing over time to evade detective approaches that rely on volume-based features, as we have seen in Section  \ref{sec:fallback_channel}, which are been quite a popular approach in the past two decades.

The multi-level encrypted channel is also popular and is used by 54.5\% of APT campaigns to evade decryption if the TLS traffic is blocked unless the decrypt key is available for the web proxy. Therefore, even if the traffic is decrypted, another layer is still held, such as an encoding technique or block cipher, as described in Section \ref{sec:encrypted_channel}. The next popular TTP is domain fronting and multi-hop proxy, which is more than half (51.5\%) of APT campaigns exploit such technique together and 24.2\% only use domain fronting, to conceal the location of the remote \cc\ server as described in Section \ref{sec:Multi_hop_Proxy}.
Other DNS-based TTPs are also found, with 27.2\% using dynamic DNS, and 24.2\% exploiting DGA or DNS tunneling. This leads to the importance of considering the detection of malicious domains and UDP-based traffic.


\section{Conclusion}
In this paper, we highlight the importance of developing a successful APT detection strategy, which can be achieved by, first, studying the network-based TTPs. These TTPs pose a challenge when it comes to distinguishing between malicious and legitimate activities. Consequently, when formulating approaches for the next generation of NIDS against APTs, it is imperative to consider the particular context of the attack.

We analyze 33 APT traffic campaigns (Tables \ref{tab:long_table1} and \ref{tab:long_table2}) in terms of many features of TTPs, including evasion techniques, protocols, payloads, obfuscation, and channels. We observe several APT campaigns use zero-day vulnerabilities, and we denote that in Table \ref{tab:long_table1}. For instance, Stuxnet uses four zero-day exploits, while Elderwood uses eight \cite{aptElderwoodA}. However, other APTs such as Taidoor reuse exploits \cite{aptElderwoodA}. These different exploits provide multiple persistence against targeted organizations. In addition to that, we discuss some evasion techniques used by a wide range of campaigns and summarized in Table \ref{table:2.2}.  

We conclude that typical malware or multi-stage attacks are run normally by individuals or small teams. On the other hand, APTs are launched by a group of highly skilled people and mostly funded by governments. 
Finally, we define the most popular network-based TTPs used by APTs. We focused on HTTP(S) and DNS protocols and categorized 13 TTPs related to these protocols since HTTP(S) and DNS are the popular protocols among APTs with 81\% and 45\%, respectively. We present several examples based on our datasets \cite{alageel2021hawkeye,EarlyCrowRepo} in addition to further resources from the industry.


\end{document}